\documentclass[twocolumn,prb,superscriptaddress,longbibliography]{revtex4-2}
\pdfoutput=1
\usepackage{graphicx}
\usepackage{color}
\usepackage{amsmath}
\usepackage{enumitem}
\usepackage{amssymb}
\usepackage[normalem]{ulem}
\usepackage{adjustbox}
\usepackage{bm}
\usepackage{braket}
\usepackage{physics}
\usepackage{dcolumn}
\usepackage{hyperref}
\usepackage{siunitx}
\usepackage{pdfpages}
\makeatletter
\patchcmd{\@outputpage@head}{\@ifx{\LS@rot\@undefined}{}{\LS@rot}}{}{}{}
\makeatother

\hypersetup{
	colorlinks = true,
	linkcolor = blue,
	citecolor = blue,
	urlcolor  = blue,
}

\begin{document}
	
	\title{Theory of spin-wave transport in ferromagnet-superconductor heterostructures: Negative refraction, perfect imaging and temperature-controlled spin-wave optics}

	\author{Tomas T. Osterholt}
	\affiliation{%
		Institute for Theoretical Physics, Utrecht University, 3584CC Utrecht, The Netherlands\\
	}%

    \author{Thijs van der Meer}
	\affiliation{%
		Institute for Theoretical Physics, Utrecht University, 3584CC Utrecht, The Netherlands\\
	}%
	
	\author{Pim H. Vree}
	\affiliation{%
		Kavli Institute of Nanoscience, Delft University of Technology, 2628 CJ Delft, The Netherlands\\
	}%
	
	\author{Merel A. Bouma}
	\affiliation{%
		Kavli Institute of Nanoscience, Delft University of Technology, 2628 CJ Delft, The Netherlands\\
	}

    \author{Michael Borst}
	\affiliation{%
		Kavli Institute of Nanoscience, Delft University of Technology, 2628 CJ Delft, The Netherlands\\
	}

	\author{Toeno van der Sar}
	\affiliation{%
		Kavli Institute of Nanoscience, Delft University of Technology, 2628 CJ Delft, The Netherlands\\
	}

	\author{Rembert A. Duine}
	\affiliation{%
		Institute for Theoretical Physics, Utrecht University, 3584CC Utrecht, The Netherlands\\
	}
	\affiliation{Department of Applied Physics, Eindhoven University of Technology,
		P.O. Box 513, 5600 MB Eindhoven, The Netherlands
	}%
	
	\date{\today}
	
	\begin{abstract}
		We investigate spin-wave transport in ferromagnetic insulator–superconductor (FMI-SC) heterostructures and develop a general theoretical framework for spin-wave optics in these hybrid systems. We demonstrate that Meissner screening by the superconductor gives rise to a range of unconventional wave phenomena, including negative phase- and group-velocity refraction, and reflection and refraction laws that differ fundamentally from their optical counterparts. Within this framework, we derive the spin-wave Fresnel equations governing reflection and transmission at FMI-SC interfaces and show that the scattering properties exhibit a pronounced temperature dependence, enabling tunable spin-wave mirrors and refractive elements. Most strikingly, we find that superconducting screening can produce nearly straight isofrequency contours, far flatter than the kinked, intrinsically curved contours attainable in conventional dipolar spin-wave systems. We show that these straight contours enable functionalities such as perfect spin-wave imaging, efficient waveguiding, and interferometric elements, such as phase shifters and beam splitters, with unconventional properties. Our results establish FMI–SC heterostructures as a versatile platform for temperature-tunable spin-wave optics and interferometric magnonic devices.
	\end{abstract}
	
	\maketitle
	
	\section{Introduction}
Spintronics exploits the spin degree of freedom of electrons to develop novel approaches for information storage, processing, and communication \cite{Wolf2001,Zutic2004,Hoffmann2015,Hirohata2020}. Among the various spin excitations, spin waves (or magnons) have attracted particular attention because they enable the transport of information through collective magnetic excitations rather than dissipative charge currents. The low energy consumption and rich dynamical properties of spin waves make them promising candidates for future information technologies, ranging from spin-based logic devices to unconventional computing architectures \cite{Chumak2015,Pirro2021,Yuan2022,Csaba2017,Candido2020,Yuan2023}.

The wave nature of spin excitations provides a natural connection between magnonics and the broader field of wave-based information processing. Similar to photonic systems, spin waves can exhibit effects such as interference \cite{Mukherjee2012,Rousseau2015,Kanazawa2016,Papp2021}, diffraction \cite{Loayza2018,Vlaminck2023,Wang2024} and refraction \cite{Stigloher2016,Hioki2020,Mieszczak2020}, which can be exploited for information processing beyond conventional semiconductor-based electronics. In this context, spin-wave optics aims to manipulate magnons as information carriers through the controlled propagation and transformation of wave signals. The aforementioned wave phenomena, together with fundamental spin-wave properties such as group velocity and polarization, are governed by the magnon dispersion relation. Consequently, the ability to engineer this dispersion is essential for controlling spin-wave propagation and realizing functional magnonic devices. One approach to dispersion control that has been explored is the modification of intrinsic magnetic interactions within the material, which can be achieved, for example, through layer twisting in few-layer magnetic systems \cite{Li2020,Wang2023,Oriekhov2025,Osterholt2025} or by applying strain \cite{Weiler2011,Liu2022}. Other approaches include varying the magnetic layer thickness \cite{Demokritov2001,Tacchi2015,Qin2020}, which modifies the relative contributions of exchange and dipolar interactions, and applying external fields, including electric fields, which can influence the dispersion through mechanisms such as multiferroic coupling \cite{Rovillain2010,Zhu2017,Qin2019}.

In recent years, ferromagnet-superconductor heterostructures have also attracted interest as a way to modify spin-wave properties through superconducting proximity effects \cite{Golovchanskiy2018,Golovchanskiy2020,Yu2022,Borst2023,Ghirri2024,Yu2026}. The underlying mechanism is the interaction between the superconductor and the stray dipolar fields generated by spin waves in the ferromagnet. Through the Meissner effect, the superconductor screens these fields and generates a counteracting magnetic field, which in turn leads to a modification of the spin-wave dispersion.

The ability of superconductors to influence spin-wave properties has led to several interesting proposals and observations. Theoretical work has shown that superconducting regions can act as magnonic gates by strongly reflecting spin waves through Meissner screening, enabling functionalities such as magnon confinement \cite{Yu2022}. More recently, it was predicted that the interaction between spin waves and superconducting screening currents can even enhance magnon transport by significantly increasing the spin-wave group velocity \cite{Zhou2024}. These predictions have been followed by experimental observations of hybrid spin-wave–Meissner-current modes \cite{Borst2023}, where nitrogen-vacancy magnetometry \cite{Balasubramanian2008,Taylor2008,Casola2018,Osterholt2024} revealed strong modifications of the spin-wave wavelength in ferromagnet–superconductor heterostructures, with a pronounced dependence on temperature due to the superconducting transition.

These and other theoretical proposals and experimental observations demonstrate the potential of ferromagnet–superconductor heterostructures as a platform for magnonic applications. The strong temperature dependence arising from the superconducting transition, combined with the wave nature shared by optical and spin-wave systems, makes these structures particularly interesting for realizing temperature-tunable spin-wave devices. In this context, spin-wave optics aims to manipulate magnons in analogy with photons by exploiting wave phenomena such as reflection, refraction, interference, and diffraction. The development of functional spin-wave optical elements therefore requires a detailed understanding of how spin waves are reflected and transmitted at ferromagnet–superconductor interfaces. 

In this work, we address this question by developing a rigorous description of the reflection and transmission properties of ferromagnet–superconductor heterostructures. In particular, we derive the laws governing reflection and refraction of dipolar spin waves in these systems and demonstrate their strong dependence on temperature. We further obtain the corresponding reflection and transmission coefficients, analogous to the Fresnel coefficients in optics. Using this framework, we also identify several unconventional wave phenomena enabled by superconducting Meissner screening. Most notably, we demonstrate negative phase- and group-velocity refraction and reflection, as well as the emergence of nearly straight isofrequency contours that are considerably flatter than those achievable in conventional dipolar spin-wave systems. We highlight several functionalities enabled by these straight isofrequency contours, including perfect spin-wave imaging, efficient waveguiding, and temperature-tunable interferometric elements such as phase shifters and beam splitters with unconventional properties.

The remainder of this paper is organized as follows. In Section \ref{Section Dispersion Relation}, we derive the dispersion relation of dipolar spin waves in ferromagnetic insulator–superconductor (FMI–SC) heterostructures. In Section \ref{Section Reflection Refraction}, we determine the temperature-dependent reflection and refraction properties of the system and demonstrate the emergence of nearly flat isofrequency contours. Finally, in Section \ref{Section Applications}, we discuss the range of spin-wave functionalities enabled by the unique properties of FMI–SC heterostructures.

\section{Dispersion Relation}\label{Section Dispersion Relation}
\begin{figure*}[t]
    \centering
    \includegraphics[width=0.9\linewidth]{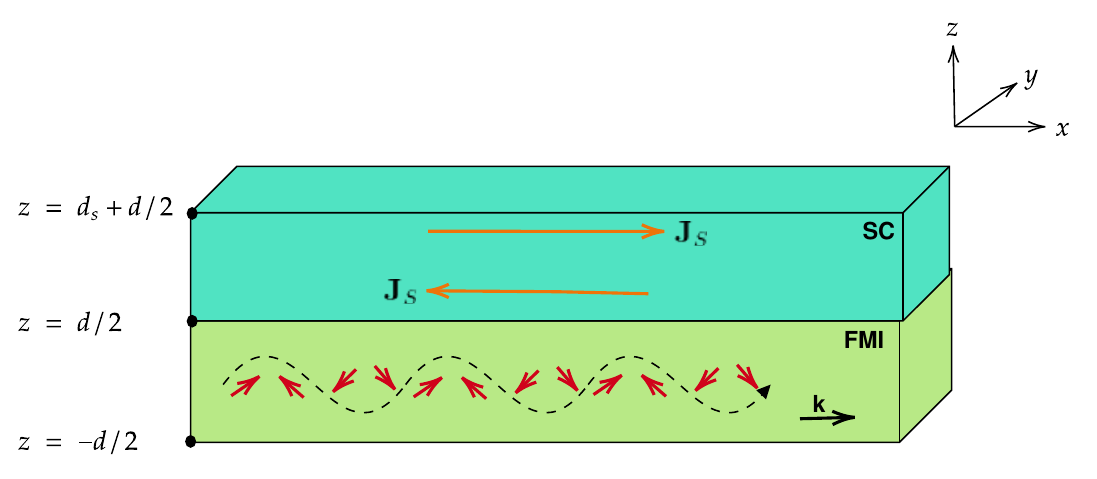}
    \caption{\textbf{FMI-SC heterostructure.} A spin wave with wavevector $\mathbf{k}$ is traveling through the ferromagnetic insulator (FMI), resulting in Meissner screening from the superconductor (SC) via the generation of a supercurrent density $\mathbf{J}_S$.}
    \label{Figure FMI-SC}
\end{figure*}

  Here, we derive the dispersion relation $\omega(\mathbf{k})$ for a spin wave with wavevector $\mathbf{k} = k_x \hat{\mathbf{x}} + k_y \hat{\mathbf{y}}$ in an FMI-SC heterostructure. The ferromagnetic insulator (FMI) is centered at $z = 0$ and has a finite thickness $d$ along the $z$-direction, while the superconducting film has a finite thickness $d_s$ and occupies the region between $z = d/2$ and $z = d/2 + d_s$. Both the superconductor and the FMI are assumed to extend to infinity in the $x$- and $y$-directions. We further assume that a static magnetic field $\mathbf{B}_0 = B_0 \,\hat{\mathbf{y}}$ is applied, resulting in a saturation magnetization $\mathbf{M}_S = M_S\, \hat{\mathbf{y}}$ in the FMI. Considering a sufficiently thin magnetic film ($kd \ll 1$), the spin-wave magnetization density $\mathbf{m}_{\mathbf{k}}(\mathbf{r},t)$ inside the film can then be considered to be uniform along the $z$-direction and, consequently, it takes the form
    \begin{align}\label{Equation Spin-Wave Magnetization}
    \mathbf{m}_{\mathbf{k}}(\mathbf{r},t)
    = \big[\theta(z+d/2)-\theta(z-d/2)\big]
    e^{i (\mathbf{k} \cdot \boldsymbol{\rho}-\omega t)}
    \begin{pmatrix}
        i m_x\\
        0\\
        m_z
        \end{pmatrix}
    .
    \end{align}
Here, $\boldsymbol{\rho} = x \hat{\mathbf{x}} + y \hat{\mathbf{y}}$, $\theta(z)$ denotes the Heaviside step function, $t$ is time, and $m_x,m_z \in \mathbb{C}$ are coefficients corresponding to the amplitudes of the magnetization-density components. An illustration of the system is given in Fig \ref{Figure FMI-SC}.

The derivation of the dispersion relation $\omega(\mathbf{k})$ proceeds in two steps. The spin wave described by Eq.~\eqref{Equation Spin-Wave Magnetization} generates a dipolar magnetic field $\mathbf{B}_d(\mathbf{r},t)$, which in turn results, through the Meissner effect, in the generation of an additional magnetic field $\mathbf{B}_{\mathrm{sc}}(\mathbf{r},t)$ generated by the superconductor. We refer to this induced field as the response field. Our first step therefore consists of determining $\mathbf{B}_{\mathrm{sc}}(\mathbf{r},t)$ using classical electromagnetic theory. Finally, once the response field has been obtained, we use it to derive $\omega(\mathbf{k})$ via the Landau-Lifshitz equation.

It should be mentioned that the dispersion relation has been derived earlier in Ref.~\cite{Borst2023}, which includes several of the authors of this paper, building on the formalism developed in Ref.~\cite{Yu2022}. Here, we present a more general approach for calculating the response field, which, in contrast to the previous study, determines all components of $\mathbf{B}_{\mathrm{sc}}(\mathbf{r},t)$ inside the superconductor and yields a self-consistent equation applicable to arbitrary geometries beyond the specific case considered there. Nevertheless, given that the same dispersion relation $\omega(\mathbf{k})$ is obtained as in Ref.~\cite{Borst2023}, readers familiar with said work may wish to skip ahead to the next section.
	
\subsection{Derivation of the magnetic field}
To calculate the response field $\mathbf{B}_{\mathrm{sc}}(\mathbf{r},t)$ of the superconductor, we first note that the spin-wave magnetization of Eq.~\eqref{Equation Spin-Wave Magnetization} generates a dipolar magnetic field $\mathbf{B}_d(\mathbf{r},t)$, which, as shown in Appendix \ref{Appendix Integral Identities}, in the region $z>d/2$ is given by
\begin{align}\label{Equation Spin-Wave Magnetic Field}
		\mathbf{B}_{\mathrm{d}}(\mathbf{r},t) &= B_{\mathrm{d}}(\mathbf{k}) e^{i (\mathbf{k}\cdot \boldsymbol{\rho}-\omega t)}e^{-kz} 
		\begin{pmatrix}
			i k_x/k\\
			i k_y/k\\
			-1
		\end{pmatrix}
        ,
	\end{align}
    with
    \begin{align}
        B_{\mathrm{d}}(\mathbf{k}) &= - \mu_0\sinh \left( \frac{k d}{2} \right) \left[ \frac{k_x}{k} m_x +  m_z \right].
    \end{align}
    Here, $\mu_0$ denotes the vacuum permeability. Determining $\mathbf{B}_{\mathrm{sc}}(\mathbf{r},t)$ is equivalent to solving the classical electromagnetic problem of an infinite superconducting slab coupled to the dipolar magnetic field generated by the spin wave, Eq.~\eqref{Equation Spin-Wave Magnetic Field}. To this end, we decompose the total magnetic field according to its physical origin as
    \begin{align}
        \mathbf{B}(\mathbf{r},t)=\mathbf{B}_d(\mathbf{r},t)+\mathbf{B}_{\mathrm{sc}}(\mathbf{r},t),
    \end{align}
where $\mathbf{B}_d$ denotes the dipolar field generated by the spin-wave magnetization and $\mathbf{B}_{\mathrm{sc}}$ denotes the magnetic field generated by the superconducting screening currents. We stress that this decomposition does not distinguish between external and induced contributions. Rather, it is a separation according to the source of the magnetic field. In particular, $\mathbf{B}_d$ may itself depend on the total magnetic field through the spin-wave dynamics and therefore incorporates the back-action of the superconductor on the magnetic subsystem. Below, we derive a self-consistent equation for the total magnetic field $\mathbf{B}$, from which $\mathbf{B}_{\mathrm{sc}}$ can be determined explicitly.

\subsubsection{Self-consistent equation}\label{Section Self-Consistent Equation}
	As our starting point, we consider Maxwell's equations,
	\begin{align}
		\begin{cases}
			\nabla \cdot \mathbf{E} &= \dfrac{\rho_e}{\epsilon_0},\\[6pt]
			\nabla \cdot \mathbf{B} &= 0,\\[6pt]
			\nabla \times \mathbf{E} &= - \dfrac{\partial \mathbf{B}}{\partial t} ,\\[6pt]
			\nabla \times \mathbf{B} &= \mu_0\bigg( \mathbf{J}+\epsilon_0 \dfrac{\partial \mathbf{E}}{\partial t} \bigg),
		\end{cases}
	\end{align}
	with $\mathbf{E}$ the total electric field, $\rho_e$ and $\mathbf{J}$ the charge and current densities, and $\epsilon_0$ the vacuum permittivity, in combination with the London equations,
	\begin{align}
		\frac{\partial \mathbf{J}_S}{\partial t} &= \frac{1}{\mu_0\lambda_L^2} \mathbf{E}_S,\\
		\nabla \times \mathbf{J}_S &= - \frac{1}{\mu_0\lambda_L^2} \mathbf{B}_S,
	\end{align}
	with $\lambda_L$ the London penetration depth and with $\mathbf{J}_S$, $\mathbf{E}_S$ and $\mathbf{B}_S$ the supercurrent density, the electric field and magnetic field inside the superconducting region, respectively.
	
	Now, the total magnetic field in the system is given by $\mathbf{B} = \mathbf{B}_{\mathrm{d}}+\mathbf{B}_{\mathrm{sc}}$. Since $\nabla \cdot \mathbf{B} = 0$ and $\nabla \cdot \mathbf{B}_{\mathrm{d}} = 0$, we can immediately conclude that $\nabla \cdot \mathbf{B}_{\mathrm{sc}} = 0$ everywhere. This directly implies that we can derive $\mathbf{B}_{\mathrm{sc}}(\mathbf{r},t)$ from a vector potential $\mathbf{A}_{\mathrm{sc}}(\mathbf{r},t)$, such that $\mathbf{B}_{\mathrm{sc}}(\mathbf{r},t) = \nabla \times \mathbf{A}_{\mathrm{sc}}(\mathbf{r},t)$ , where
	\begin{align}
		\mathbf{A}_{\mathrm{sc}}(\mathbf{r},t) &= \frac{1}{4\pi} \int_{\mathbb{R}^3} d^3 \mathbf{r}'\frac{\nabla'\times\mathbf{B}_{\mathrm{sc}}(\mathbf{r}',t)}{|\mathbf{r}-\mathbf{r}'|}.
	\end{align}
	To determine the curl of the response field, we make use of the quasi-static approximation, which allows us to ignore the effects coming from the displacement currents. With this approximation, we immediately find that 
    \begin{align}
        \nabla \times \mathbf{B}_{\mathrm{sc}}(\mathbf{r},t) = 
        \begin{cases}
            \mu_0 \mathbf{J}_S(\mathbf{r},t), \hspace{0.1cm} \text{for} \hspace{0.1cm} \mathbf{r} \in \Omega_S,\\
             \mathbf{0}, \hspace{1.5cm} \text{for} \hspace{0.1cm} \mathbf{r} \notin \Omega_S,
        \end{cases}
    \end{align}
	where $\Omega_S$ is the region in $\mathbb{R}^3$ occupied by the superconductor. Consequently, we have
	\begin{align}\label{Equation Magnetic Potential Response Field}
		\mathbf{A}_{\mathrm{sc}}(\mathbf{r},t) &= \frac{\mu_0}{4\pi} \int_{\Omega_S} d^3 \mathbf{r}'\frac{\mathbf{J}_S(\mathbf{r}',t)}{|\mathbf{r}-\mathbf{r}'|}.
	\end{align}
	
	From the curl identity $\nabla \times (\Psi\mathbf{V}) = \Psi (\nabla \times\mathbf{V}) + (\nabla \Psi) \times \mathbf{V}$, and the gradient identity
	\begin{align}
		\nabla \biggr( \frac{1}{|\mathbf{r}-\mathbf{r}'|} \biggr) &= - \nabla'\biggr( \frac{1}{|\mathbf{r}-\mathbf{r}'|} \biggr),
	\end{align}
	we can then use Eq.~\eqref{Equation Magnetic Potential Response Field} to express the response field in the following way,
	\begin{align}
		\mathbf{B}_{\mathrm{sc}}(\mathbf{r},t) &= -\frac{\mu_0}{4\pi} \int_{\Omega_S} d^3 \mathbf{r}' \,\nabla'\biggr( \frac{1}{|\mathbf{r}-\mathbf{r}'|}\biggr)\times \mathbf{J}_S(\mathbf{r}',t).
	\end{align}
	Again using the vector identity given above, now with respect to the primed coordinates, and using the integral identity
	\begin{align}
		\int_{\Omega} d^3 \mathbf{r}(\nabla \times \mathbf{V})  &= - \oint_{\partial \Omega} d^2 \mathbf{r}(\mathbf{V}\times \hat{\mathbf{n}}) ,
	\end{align}
	with $\partial \Omega$ the boundary surface of the volume $\Omega$ and $\hat{\mathbf{n}}$ the outward normal vector to said surface, we then have
	\begin{align}
		\mathbf{B}_{\mathrm{sc}}(\mathbf{r},t) &= \frac{\mu_0}{4\pi} \int_{\Omega_S} d^3 \mathbf{r}'\frac{\nabla'\times\mathbf{J}_S(\mathbf{r}',t)}{|\mathbf{r}-\mathbf{r}'|} \nonumber \\&+ \frac{\mu_0}{4\pi} \oint_{\partial \Omega_S} d^2 \mathbf{r}' \frac{\mathbf{J}_S(\mathbf{r}',t)\times \hat{\mathbf{n}}}{|\mathbf{r}-\mathbf{r}'|}. 
	\end{align}
	Finally, using the second London equation, we arrive at the following self-consistent equation for the total magnetic field $\mathbf{B}(\mathbf{r},t)$
	\begin{align}
		\mathbf{B}(\mathbf{r},t) &= \mathbf{B}_{\mathrm{d}}(\mathbf{r},t)- \frac{1}{4\pi \lambda_L^2} \int_{\Omega_S} d^3 \mathbf{r}'\biggr[\frac{\mathbf{B}(\mathbf{r}',t)}{|\mathbf{r}-\mathbf{r}'|}\biggr] \nonumber\\&+ \frac{1}{4\pi} \oint_{\partial \Omega_S} d^2 \mathbf{r}' \frac{(\nabla' \times \mathbf{B}(\mathbf{r}',t))\times \hat{\mathbf{n}}}{|\mathbf{r}-\mathbf{r}'|}. \label{Equation Self-Consistent Equation}
	\end{align}
    It should be noted that, at this stage of the derivation, the specific geometry of the superconducting region $\Omega_S$ has not entered the calculation, apart from the requirement that its boundary $\partial\Omega_S$ forms a closed surface. Consequently, the self-consistent equation derived above is valid for arbitrary geometries of $\Omega_S$ satisfying this condition.

	\subsubsection{Solution inside superconductor}\label{Section Solution to Self-Consistent Equation in Superconductor}
	To solve the self-consistent equation, Eq.~\eqref{Equation Self-Consistent Equation}, we note that it is sufficient to determine $\mathbf{B}(\mathbf{r},t)$ inside the superconducting region $\Omega_S$. Using the symmetry of the system, we expect a reasonable ansatz for the magnetic field to be of the following form,
	\begin{align}
		\mathbf{B}(\mathbf{r},t) &= e^{i(\mathbf{k} \cdot \boldsymbol{\rho}-\omega t)}
		\begin{pmatrix}
			B_x(z)\\
			B_y(z)\\
			B_z(z)
		\end{pmatrix}
		.
	\end{align}
	Since we require that $\nabla \cdot \mathbf{B} = 0$ everywhere, the magnetic field components $B_x(z)$, $B_y(z)$ and $B_z(z)$ have to satisfy the equation
	\begin{align}
		i k_x B_x(z) + i k_y B_y(z) + \frac{d B_z(z)}{d z} = 0.
	\end{align}
	Given that the applied spin wave field is of the form $e^{i(\mathbf{k} \cdot \boldsymbol{\rho}-\omega t)} e^{-k|z|} (i k_x, i k_y, -k)^T$, we propose that
	\begin{align}
		B_x(z) &= i k_x f_x(z),\\
		B_y(z) &= i k_y f_y(z),
	\end{align}
	for some yet-to-be-determined functions $f_x(z)$ and $f_y(z)$. Because of the in-plane symmetry of the setup, we expect $f_x(z) = f_y(z) \equiv f(z)$, which yields the following ordinary differential equation,
	\begin{align}
		\frac{d B_z(z)}{d z} = k^2 f(z).
	\end{align}
	To find $f(z)$ inside the superconducting region, we take inspiration from the solution for an infinite superconducting slab in a parallel and uniform applied field, and we therefore make the ansatz
	\begin{align}
		f(z) &=C_1 e^{-\zeta z} + C_2 e^{\zeta z}, \hspace{0.1cm} \text{for} \hspace{0.1cm} z \in [d/2,d/2+d_s],
	\end{align}
	where $\zeta\in \mathbb{R}_{>0}$ and $C_1,C_2 \in \mathbb{C}$ are constants to be determined. Consequently, one finds that
	\begin{align}
		B_z(z) &=
		- \frac{k^2}{\zeta} C_1 e^{-\zeta z} + \frac{k^2}{\zeta} C_2 e^{\zeta z} + C_3,
	\end{align}
	for $z \in [d/2,d/2+d_s]$, with $C_3 \in \mathbb{C}$ also constant. 
	
	As shown in Appendix \ref{Appendix Derivation of C, and kappa Coefficients}, these constants can be determined by plugging the ansatz into Eq.~\eqref{Equation Self-Consistent Equation}, after which we find the following solution:
	\begin{align}
		\left\{
		\begin{aligned}
			\zeta &= \sqrt{k^2+\frac{1}{\lambda_L^2}},\\
			C_1 &= \frac{2 \zeta \, B_{\mathrm{d}}(\mathbf{k}) e^{(\zeta-k)d/2}}{a_{k,+}k^2}\frac{1}{1-\big(\frac{a_{k,-}}{a_{k,+}}\big)^2 e^{-2 \zeta d_s}},\\
			C_2&= \frac{2 a_{k,-}\zeta \, B_{\mathrm{d}}(\mathbf{k})e^{-(\zeta+k)d/2}}{a_{k,+}^2k^2}\frac{e^{-2 \zeta d_s}}{1-\big(\frac{a_{k,-}}{a_{k,+}}\big)^2 e^{-2 \zeta d_s}},\\
			C_3  &= 0,
		\end{aligned}
		\right.
	\end{align}
	where
	\begin{align}
		a_{k,\pm} &= 1 \pm \zeta/k.
	\end{align}
	
	\subsubsection{Solution inside ferromagnetic insulator}\label{Section Solution inside magnetic insulator}
	Having determined $\mathbf{B}(\mathbf{r},t)$ inside the superconducting region, we can immediately obtain its solution inside the FMI via Eq.~\eqref{Equation Self-Consistent Equation}. We find that the response field inside the FMI is given by
	\begin{align}
		\mathbf{B}_{\mathrm{sc}}^{\mathrm{in}}(\mathbf{r},t) &= B_{\mathrm{sc}}^{\mathrm{in}}(\mathbf{k}) e^{i(\mathbf{k} \cdot \boldsymbol{\rho}-\omega t)} e^{kz}
		\begin{pmatrix}
			i k_x/k\\i k_y/k \\ 1
		\end{pmatrix}
		,
	\end{align}
	with
	\begin{align}
		B_{\mathrm{sc}}^{\mathrm{in}}(\mathbf{k})
		&=  \dfrac{B_d(\mathbf{k}) e^{-k d}}{k^2 \lambda_L^2} \biggr( \frac{1-e^{-2\zeta d_s}}{a_{k,+}^2 - a_{k,-}^2 e^{-2\zeta d_s}}  \biggr). 
	\end{align}
	Here, we have included the superscript `in' to stress that we are considering the value of the response field \textit{inside} the FMI. Now, we emphasized earlier that the magnetization density inside the magnetic film can be considered to be uniform along the $z$-direction in the thin-film limit, which is valid when $k d \ll 1$. Working in said limit, internal consistency requires us to work with fields averaged over the film along $z$. The average response field inside the FMI can be calculated straightforwardly, and we have
	\begin{align}
		\mathbf{B}_{\mathrm{sc},\mathrm{ave}}^{\mathrm{in}}(\boldsymbol{\rho},t) &= \frac{1}{d} \int_{-d/2}^{d/2} dz\, \mathbf{B}_{\mathrm{sc}}^{\mathrm{in}}(\mathbf{r},t)\nonumber\\&=	B_{\mathrm{sc},\mathrm{ave}}^{\mathrm{in}}(\mathbf{k}) e^{i(\mathbf{k} \cdot \boldsymbol{\rho}-\omega t)} 
		\begin{pmatrix}
			i k_x/k\\i k_y/k \\ 1
		\end{pmatrix}
		,
	\end{align}
	with
	\begin{align}
		B_{\mathrm{sc},\mathrm{ave}}^{\mathrm{in}}(\mathbf{k}) &= - \frac{\omega_{\mathrm{sc}}(k)}{\gamma  M_S} \biggr[ \frac{k_x}{k} m_x + m_z \biggr].
	\end{align}
	For future convenience, we have introduced the superconductor-response frequency $\omega_{\mathrm{sc}}(k)$, which is given by
	\begin{align}
		\omega_{\mathrm{sc}} (k) &= \frac{\gamma\mu_0 M_S\, g^2(k)\, d}{2 k \lambda_L^2} \frac{1-e^{-2 \zeta d_s}}{a_{k,+}^2-a_{k,-}^2 e^{-2\zeta d_s}},
	\end{align}
	with $\gamma$ and $M_S$ the gyromagnetic ratio and the saturation magnetization of the insulator, respectively, and where
	\begin{align}
		g(k) &= \frac{1}{k d} \bigg(1-e^{-kd} \bigg).
	\end{align}
    In a similar fashion, the average dipolar 
    demagnetizing field generated by the spin wave is given by
    \begin{align}
        \mathbf{H}^{\mathrm{in}}_{\mathrm{d},\mathrm{ave}}(\boldsymbol{\rho},t) &= \dfrac{e^{i(\mathbf{k}\cdot \boldsymbol{\rho}-\omega t)}}{k^2}
        \begin{pmatrix}
            i k_x^2 [g(k)-1]m_x\\
            i k_x k_y[g(k)-1]m_x\\
            -k^2 g(k) m_z
        \end{pmatrix}
        .
    \end{align}
    A derivation of this equation is provided in Appendix \ref{Appendix Integral Identities}.
	
	\subsection{Landau-Lifshitz equation}
	
	\subsubsection{Derivation of dispersion relation}
	To determine the dispersion relation of the spin waves in the system, we consider the Landau-Lifshitz equation in the absence of Gilbert damping. Denoting with $\mathbf{M}(\mathbf{r},t) = \mathbf{M}_S + \mathbf{m}_{\mathbf{k}}(\mathbf{r},t)$ the total magnetization density inside the FMI, where $\mathbf{M}_S$ is the static, uniform (saturation) magnetization density and $\mathbf{m}_{\mathbf{k}}(\mathbf{r},t)$ is the spin-wave magnetization density, this equation is given by
	\begin{align}
		\frac{d \mathbf{M}}{d t} &= - \gamma \mu_0 \mathbf{M} \times \mathbf{H}_{\mathrm{eff}},
	\end{align}
	with $\mathbf{H}_{\mathrm{eff}}$ the effective field. This effective field is calculated by taking the functional derivative of the magnetic free energy $E[\mathbf{M}(\mathbf{r},t)]$ with respect to the magnetization density,
	\begin{align}
		\mathbf{H}_{\mathrm{eff}}(\mathbf{r},t) = - \frac{1}{\mu_0} \frac{\delta E[\mathbf{M}(\mathbf{r},t)]}{\delta \mathbf{M}(\mathbf{r},t)}.
	\end{align}
	For our system, we consider four distinct contributions to the effective field. The first contribution is due to the exchange field $\mathbf{H}_{\mathrm{exc}}(\mathbf{r},t) = A \nabla^2 \mathbf{m}_{\mathbf{k}}(\mathbf{r},t)$ inside the insulator, with $A$ the exchange stiffness. The second contribution is due to the static and uniform applied magnetic field $\mathbf{B}_0 = B_0 \hat{\mathbf{y}}$, which also gives us $\mathbf{M}_S = M_S \hat{\mathbf{y}}$, with $B_0,M_S > 0$. Since the direction of the applied field is in the plane of the magnetic film, we have a static and uniform contribution $\mathbf{H}_0 = B_0/\mu_0\hat{\mathbf{y}}$ to the effective field \footnote{For an infinite rectangular superconducting film in a static and uniform applied field $\mathbf{B}_0$, the field outside the superconductor will always be equal to $\mathbf{B}_0$ \cite{Tinkham1996}. One therefore does not have to consider a back-reaction from the superconductor to this applied field.}. The last two contributions to consider are the dipolar field $\mathbf{H}_{\mathrm{d},\mathrm{ave}}^{\mathrm{in}}$ generated by the spin waves and the response-field contribution $\mathbf{H}_{\mathrm{sc},\mathrm{ave}}^{\mathrm{in}} = \mathbf{B}_{\mathrm{sc},\mathrm{ave}}^{\mathrm{in}}/\mu_0$ \footnote{All the effects of the response field on the magnetization of the film are included in $\mathbf{H}^{\mathrm{in}}_{\mathrm{d},\mathrm{ave}}$, and we therefore have $\mathbf{H}_{\mathrm{sc},\mathrm{ave}}^{\mathrm{in}} = \mathbf{B}_{\mathrm{sc},\mathrm{ave}}^{\mathrm{in}}/\mu_0$. } of the superconductor, respectively. Collecting all these contributions,  the effective field is thus given by
	\begin{align}
		\mathbf{H}_{\mathrm{eff}} &= \mathbf{H}_0+\mathbf{h}_{\mathrm{eff}}.
	\end{align}
    where the dynamic effective field $\mathbf{h}_{\mathrm{eff}}$, resulting purely from the spin-wave magnetization density $\mathbf{m}_{\mathbf{k}}(\mathbf{r},t)$, is equal to
    \begin{align}
        \mathbf{h}_{\mathrm{eff}} = \mathbf{H}_{\mathrm{exc}}+ \mathbf{H}_{\mathrm{d},\mathrm{ave}}^{\mathrm{in}} +\mathbf{H}_{\mathrm{sc},\mathrm{ave}}^{\mathrm{in}}.
    \end{align}
	We emphasize that other contributions, such as those arising from magnetic anisotropy and the Dzyaloshinskii–Moriya interaction, can straightforwardly be included as well, but for the sake of clarity and to emphasize the effect of the superconductor-response field, we focus only on the four terms given above.
	
	Now, assuming $M_S\gg |\mathbf{m}_{\mathbf{k}}(\mathbf{r},t)|$ and keeping only terms linear in $\mathbf{m}_{\mathbf{k}}(\mathbf{r},t)$, the Landau-Lifshitz equation for the FMI-SC heterostructure becomes
	\begin{align}
		\frac{d \mathbf{m}_{\mathbf{k}}}{d t} = &- \gamma \mu_0 \mathbf{M}_S \times \mathbf{h}_{\mathrm{eff}} - \gamma B_0 \mathbf{m}_{\mathbf{k}} \times \hat{\mathbf{y}},
	\end{align}
	where we have used the fact that $\mathbf{h}_{\mathrm{eff}}$ depends linearly on the spin-wave magnetization components. From the above equation, we immediately see that
	\begin{align}
		\biggr( \frac{d \mathbf{m}_{\mathbf{k}}}{d t} \biggr)_y = 0,
	\end{align}
	and we can therefore absorb $m_y$ into $\mathbf{M}_S$  \footnote{Due to the symmetry of the system, a time-independent $m_y$ has to be spatially uniform inside the magnetic film.}. Consequently, we have
	\begin{align}
		\mathbf{m}_{\mathbf{k}}(\mathbf{r},t) &= m_x(\mathbf{r},t) \hat{\mathbf{x}} + m_z(\mathbf{r},t) \hat{\mathbf{z}}.
	\end{align}
	Now, assuming exchange boundary conditions \cite{Rado1959} along $z$ on the magnetization density, from which it follows that the solution homogeneous in $z$ is the state with the lowest energy, we expect the spin-wave magnetization density to be given by Eq.~\eqref{Equation Spin-Wave Magnetization}. Plugging this ansatz in the Landau-Lifshitz equation, we immediately find that it implies that
	\begin{align}\label{Equation Inverse Susceptibility Tensor}
		\biggr[ - i \omega\, \hat{1} + \hat{\mathcal{V}}  \biggr] 
		\begin{pmatrix}
			i m_x\\m_z
		\end{pmatrix}
		=
		\mathbf{0},
	\end{align}
	with $\hat{1}$ the $2 \times 2$ identity matrix and with $\hat{\mathcal{V}}$ given by
	\begin{align}
		\hat{\mathcal{V}} &= 
		\begin{pmatrix}
			i \omega_{\mathrm{sc}}(k) \dfrac{k_x}{k} & - \omega_1(k)-\omega_{\mathrm{sc}}(k)\\
			\omega_2(\mathbf{k})+\omega_{\mathrm{sc}}(k) \bigg(\dfrac{k_x}{k}\bigg)^2 & i \omega_{\mathrm{sc}}(k) \dfrac{k_x}{k}
		\end{pmatrix}
	\end{align}
	where we have introduced the angular frequencies $\omega_1(k)$ and $\omega_2(\mathbf{k})$,
	\begin{align}
		\omega_1(k) &= \gamma B_0+\gamma D k^2+ \gamma \mu_0 M_S\, g(k),\\
		\omega_2(\mathbf{k}) &= \gamma B_0+\gamma D k^2+ \gamma \mu_0 M_S \bigg[1-g(k)\bigg]\bigg(\dfrac{k_x}{k}\bigg)^2,
	\end{align}
	with  $D = \mu_0 M_S A$. Now, a non-trivial solution to Eq.~\eqref{Equation Inverse Susceptibility Tensor} is possible if and only if the matrix $- i \omega\, \hat{1} + \hat{\mathcal{V}}$ is not invertible, and the spin wave frequencies $\omega$ are thus determined via
	\begin{align}
		\det\biggr(- i \omega\, \hat{1} + \hat{\mathcal{V}}\biggr) = 0,
	\end{align}
	which straightforwardly yields the desired dispersion relation,
	\begin{widetext}
		\begin{align}\label{Equation Spin Wave Dispersion}
			\omega_{\pm}(\mathbf{k}) = \omega_{\mathrm{sc}}(k) \dfrac{k_x}{k} \pm \sqrt{\omega_1(k) \omega_2(\mathbf{k})} \sqrt{1+\dfrac{\omega_{\mathrm{sc}}(k)\bigg[\omega_2(\mathbf{k})+\omega_1(k) \bigg( \dfrac{k_x}{k}\bigg)^2\bigg]}{\omega_1(k) \omega_2(\mathbf{k})}+\frac{\omega_{\mathrm{sc}}^2(k) \bigg( \dfrac{k_x}{k}\bigg)^2}{\omega_1(k) \omega_2(\mathbf{k})}}.
		\end{align} 
	\end{widetext}
	In what follows, we disregard the negative frequency solution and define the dispersion relation via $\omega(\mathbf{k}) \equiv \omega_+(\mathbf{k})$. Although this dispersion relation is rather complicated, it can be simplified significantly by realizing that, in the thin-film limit,
    one typically has $\omega_{\mathrm{sc}}(k) \ll \omega_1(k), \omega_2(\mathbf{k})$. Using a Taylor-expansion up to first order in $\omega_{\mathrm{sc}}(k)$, we then arrive at the following simplified expression for the dispersion relation,
	\begin{align}
		\omega(\mathbf{k}) \approx \sqrt{\omega_1(k)\omega_2(\mathbf{k})} + \dfrac{\omega_{\mathrm{sc}}(k)}{2\eta_0(\mathbf{k})}\left(1+\eta_0(\mathbf{k})\dfrac{k_x}{k}\right)^2,
	\end{align}
	where
	\begin{align}
		\eta_0(\mathbf{k}) &= \sqrt{\dfrac{\omega_1(k)}{\omega_2(\mathbf{k})}}.
	\end{align}
    Finally, in the absence of the superconductor, or equivalently, when $\lambda_L \rightarrow \infty$, we obtain the dispersion relation $\omega_0(\mathbf{k})$ for the bare FMI,
    \begin{align}\label{Equation Dispersion Bare Insulator}
        \omega_0(\mathbf{k}) &= \sqrt{\omega_1(k) \omega_2(\mathbf{k})} .
    \end{align}

	\subsubsection{Derivation of eigenmodes}
	To determine the eigenvector $(i m_x,m_z)^T$ of Eq.~\eqref{Equation Inverse Susceptibility Tensor}, we solve the equation
	\begin{align}
		\biggr[ - i \omega\, \hat{1} + \hat{\mathcal{V}} \biggr] 
		\begin{pmatrix}
			i m_x\\m_z
		\end{pmatrix}
		=
		\mathbf{0}.
	\end{align}
	In a straightforward manner, we then find
	\begin{align}
		m_x = \eta(\mathbf{k}) m_z.
	\end{align}
	Here, the $\mathbf{k}$-dependent spin-wave polarization ratio $\eta(\mathbf{k})$ is given by
	\begin{align}\label{Equation Eta}
		\eta(\mathbf{k}) &= \dfrac{\omega_1(k)+\omega_{\mathrm{sc}}(k)}{\omega(\mathbf{k})-\omega_{\mathrm{sc}}(k) \dfrac{k_x}{k}}.
	\end{align}
    where we note that, in the limit where $\lambda_L \rightarrow \infty$, we have $\eta(\mathbf{k}) \rightarrow \eta_0(\mathbf{k})$.

    \subsubsection{Generalization to arbitrary in-plane applied fields}\label{Section Generalization to arbitrary in-plane applied fields}
    So far, we have considered the case in which the external magnetic field is oriented along the $\hat{\mathbf{y}}$ direction, $\mathbf{B}_0 = B_0 \hat{\mathbf{y}}$, and derived the corresponding spin-wave dispersion and eigenmodes. To analyze spin-wave reflection and refraction at an interface involving the FMI-SC system, it is necessary to consider arbitrary orientations of the saturation magnetization $\mathbf{M}_S$ relative to the interface. We therefore generalize our results to an arbitrary in-plane magnetic field,
    \begin{align}
        \mathbf{B}_0 &= B_0\biggr[ -\sin \psi\, \hat{\mathbf{x}} + \cos\psi\, \hat{\mathbf{y}} \biggr],
    \end{align}
    with the corresponding saturation magnetization
    \begin{align}
        \mathbf{M}_S &= M_S\biggr[ -\sin \psi\, \hat{\mathbf{x}} + \cos\psi\, \hat{\mathbf{y}} \biggr].
    \end{align}
    Here, $\psi$ denotes the saturation magnetization angle, with $\psi = 0$ corresponding to magnetization oriented along $\hat{\mathbf{y}}$. For a spin wave with wavevector $\mathbf{k} = k_x \hat{\mathbf{x}} + k_y \hat{\mathbf{y}}$, or equivalently,
    \begin{align}
        \mathbf{k} = k \big[\cos\phi\, \hat{\mathbf{x}} + \sin\phi\, \hat{\mathbf{y}}\big],
    \end{align}
    the dispersion relation retains the form of Eq.~\eqref{Equation Spin Wave Dispersion}, provided we make the substitutions $k_x \rightarrow k \cos(\phi-\psi)$ and $k_y \rightarrow k \sin(\phi-\psi)$. In a similar fashion, we find that the eigenmodes are given by
    \begin{align}
        \mathbf{m}_{\mathbf{k}}(\mathbf{r},t) 
		= m\bigg[ \theta(z+d)-\theta(z)\bigg] e^{i (\mathbf{k} \cdot \boldsymbol{\rho}-\omega t)}
		\begin{pmatrix}
			i \eta(\mathbf{k},\psi) \cos \psi\\
			i \eta(\mathbf{k},\psi) \sin \psi\\
			1
		\end{pmatrix}
        ,
    \end{align}
    where $\eta(\mathbf{k},\psi)$ is obtained from Eq.~\eqref{Equation Eta} by applying the same substitutions as mentioned above.

	\section{Spin-wave reflection and refraction}\label{Section Reflection Refraction}
\begin{figure*}[t]
        \centering
        \includegraphics[width=0.9\linewidth]{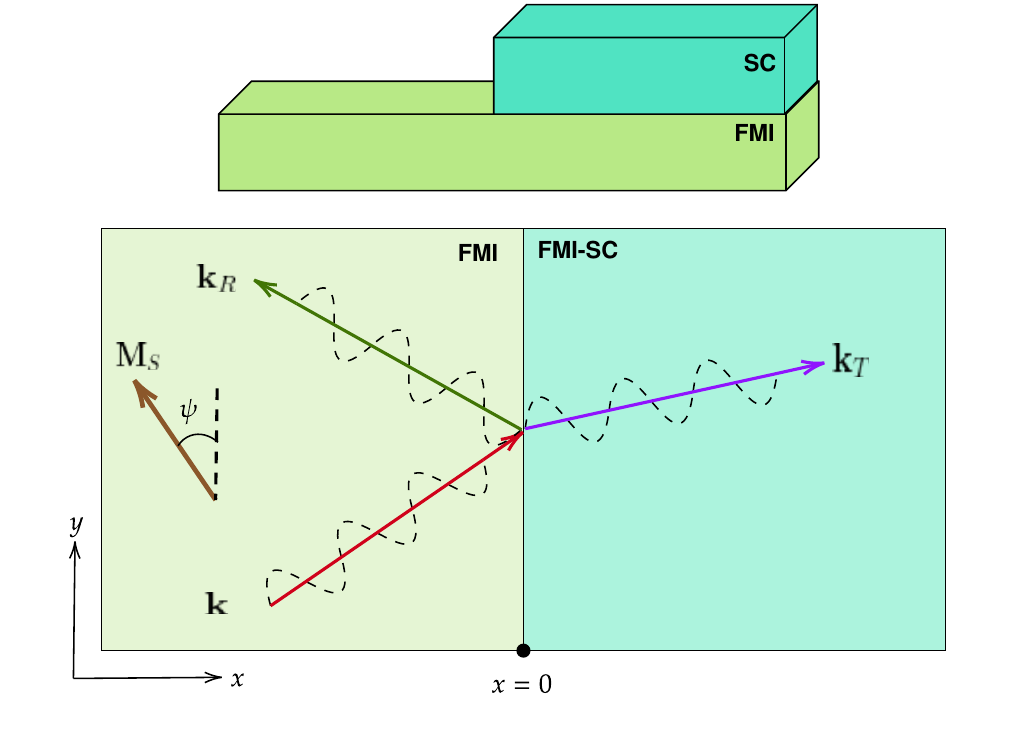}
        \caption{\textbf{Spin-wave reflection and refraction.} A spin wave with wavevector $\mathbf{k}$ is reflected and refracted at the interface between a bare FMI and an FMI-SC heterostructure located at $x=0$. Here, the wavevectors of the reflected and refracted spin waves are denoted by $\mathbf{k}_R$ and $\mathbf{k}_T$, respectively. The saturation magnetization $\mathbf{M}_S$ is oriented with an angle $\psi$ relative to the $y$-axis and is the same in both the bare FMI and the FMI-SC heterostructure}
        \label{Figure Reflection-Refraction Diagram}
    \end{figure*}

	Having derived the dispersion relation of a spin wave in a FMI-SC heterostructure, we now focus on studying spin-wave reflection and refraction at the interface between a bare FMI and said heterostructure, as illustrated in Fig.~\ref{Figure Reflection-Refraction Diagram}. Due to the complicated form of the dispersion relation, this is generally an untractable problem to tackle analytically, forcing one to employ numerical methods. Moreover, the large number of parameters entering the full dispersion relation makes it difficult, even at the numerical level, to identify which of them predominantly control the reflection and refraction processes in experimentally relevant regimes. 

	For these reasons, we will mostly concern ourselves with the dipolar regime, for which we can neglect the exchange contribution to the spin-wave frequency. In this regime, the spin-wave wavelengths are on the order of $1-100$ $\mu$m. One therefore often operates in the limit where $k d \ll 1$ and $k \lambda_L \ll 1$. Finally, if we can also assume that the thickness of the superconducting film is similar to or larger than the London penetration depth, $d_s \gtrsim \lambda_L$, the dispersion relation of Eq.~\eqref{Equation Spin Wave Dispersion}, generalized to arbitrary in-plane field angles $\psi$, can be simplified significantly, and we then have
	\begin{align}\label{Equation Approximate Dispersion Hybrid}
		\omega(\mathbf{k}) &\approx \gamma B_0 \xi + \dfrac{\gamma \mu_0 M_S}{2} k d \cos(\phi-\psi)\bigg[1+ \xi \cos(\phi-\psi) \bigg],
	\end{align}
    where the angle $\phi$ is defined relative to the interface normal via the equation $\mathbf{k} \cdot \hat{\mathbf{x}} = k \cos \phi$, and with the dipolar spin-wave \textit{ellipticity} $\xi$ given by
	\begin{align}
		\xi &= \sqrt{1+\dfrac{\mu_0 M_S}{B_0}}.
	\end{align}
    The parameter $\xi$ is called the ellipticity, because $\lim_{kd \rightarrow 0} \eta(\mathbf{k}) = \lim_{kd \rightarrow 0} \eta_0(\mathbf{k}) = \xi$ when the exchange contribution is neglected. Now, under similar approximations, we also find that the dispersion relation $\omega_0(\mathbf{k})$ for the bare FMI reduces to
	\begin{align}\label{Equation Approximate Dispersion Bare Insulator}
		\omega_0(\mathbf{k}) &\approx \gamma B_0 \xi + \dfrac{\gamma \mu_0 M_S}{4 \xi} k d \bigg[-1+ \xi^2 \cos^2(\phi-\psi) \bigg].
	\end{align}
    We will refer to the above-mentioned limit as the \textit{London dipolar limit}.

    In addition to the London dipolar limit, we will also consider the more general thin-film dipolar regime, in which the assumptions $k d \ll 1$ and negligible exchange remain valid, but the constraint $d_s \gtrsim \lambda_L$ is relaxed \footnote{It should be noted that the reflection and refraction behavior depends not only on the ratio $d_s/\lambda_L$, but also on the magnitude of the London penetration depth $\lambda_L$ relative to the spin-wave wavelength. The superconducting response is strongest in the regime $k\lambda_L \ll 1$, whereas the influence of the superconductor on spin-wave reflection and refraction becomes weak when $k\lambda_L \gtrsim 1$, even if $d_s/\lambda_L \gtrsim 1$.}. In this regime, the ratio $d_s/\lambda_L$ is allowed to take values smaller than unity, and the spin-wave dispersion therefore retains an explicit dependence on this parameter. Since the London penetration depth is itself temperature dependent through the Casimir-Gorter relation \cite{Tinkham1996}, 
        \begin{align}\label{Equation Casimir-Gorter}
    \lambda_L(T) = \dfrac{\lambda_L(0)}{\sqrt{1-\bigg( \dfrac{T}{T_c} \bigg)^4}},
    \end{align}
    where $\lambda_L(0)$ is the London penetration depth at absolute zero and $T_c$ is the critical temperature of the superconductor, the ratio $d_s/\lambda_L$ can be tuned by varying the temperature. As we will demonstrate, this enables substantial temperature control over spin-wave reflection and refraction.

    The rest of this section is organized as follows. We first discuss the properties of spin-wave reflection and refraction, combining both numerical and analytical methods of analysis. A wide variety of intriguing phenomena are observed, including both negative reflection and negative refraction, and we systematically discuss the required conditions for them to take place. Finally, we study the reflection and transmission coefficients of this system, arriving at a spin-wave analogue of the familiar Fresnel equations from optics.

\begin{figure*}[t]
    \centering
   \includegraphics[width=0.9\linewidth]{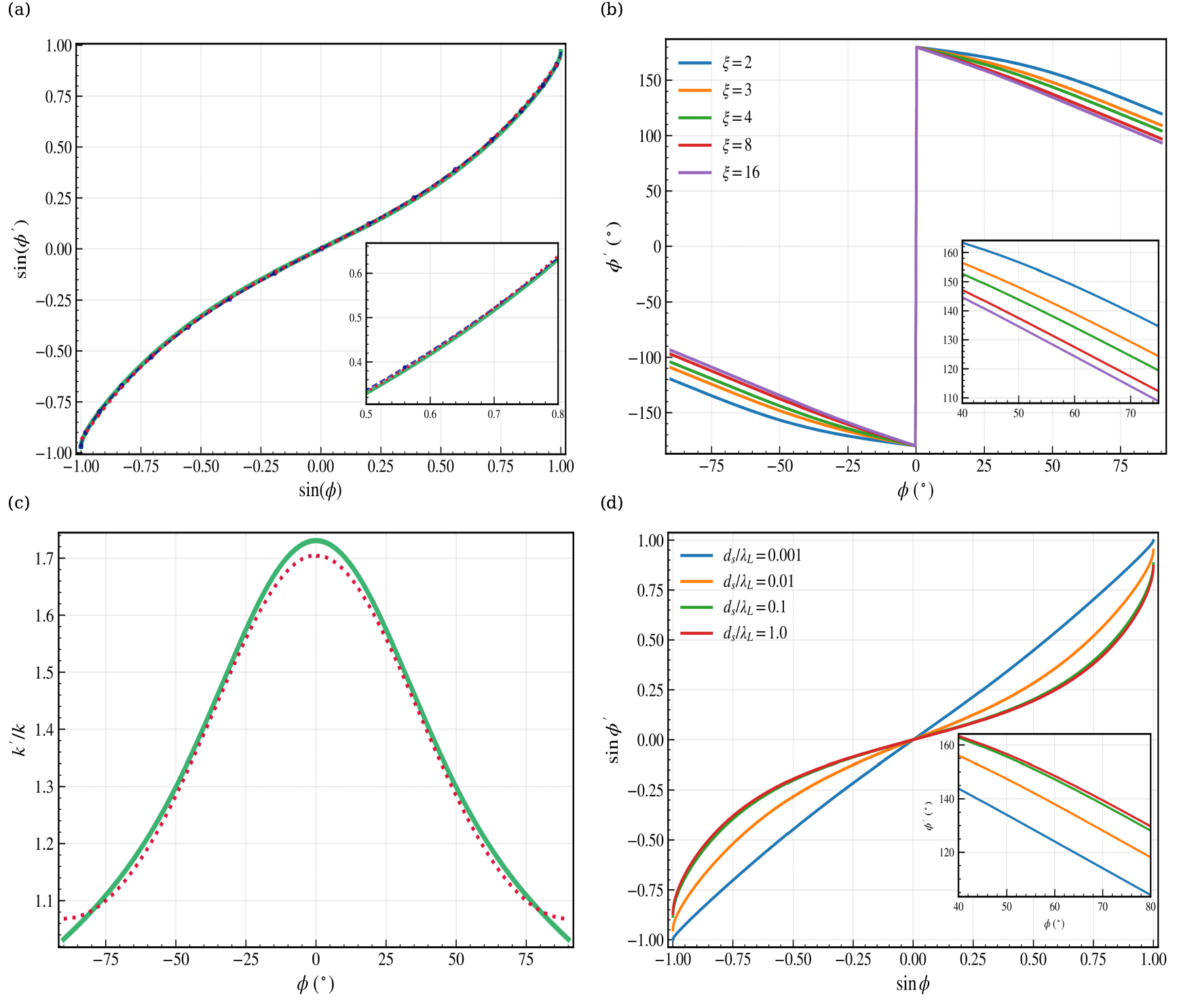}
    \caption{ \textbf{Reflection law in the FMI-SC heterostructure.} (a) Sine of the angle $\phi'$ of the reflected wavevector as a function of the sine of the angle $\phi$ of the incident wavevector. The green, solid line shows the result obtained from the exact dispersion relation, Eq.~\eqref{Equation Spin Wave Dispersion}, while the dark blue, dashed line is obtained using the polynomial approximation, Eq.~\eqref{Equation Reflection Polynomial}. The red, dotted line represents the approximate reflection law given in Eq.~\eqref{Equation Reflection Law}. (b) Reflected angle $\phi'$ as a function of the incident angle $\phi$ for different values of the ellipticity $\xi$. (c) Ratio between the magnitudes of the reflected ($k'$) and incident ($k$) wavevectors as a function of the incident angle $\phi$. The green, solid line shows the result obtained from the exact dispersion relation, Eq.~\eqref{Equation Spin Wave Dispersion}, while the red, dotted line indicates the result given by Eq.~\eqref{Equation Wavevector Magnitude Reflection}. (d) Sine of the angle $\phi'$ of the reflected wavevector as a function of the sine of the angle $\phi$ of the incident wavevector for different values of $d_s/\lambda_L$, with $d_s = 100$ nm and $\xi = 2$. In all plots, incident spin waves have wavevector magnitude $k = 2\pi/\lambda_k$ and a group velocity with a positive $x$-component. Unless stated otherwise, the following parameters were used to generate these plots: $D=0.053\,\mathrm{mT}\,\mu \mathrm{m}^2$, $\mu_0 M_S = 250\,\mathrm{mT}$, $\xi = 4$, $d = 100\,\mathrm{nm}$, $d_s = 1\,\mu\mathrm{m}$, $\lambda_L = 100\,\mathrm{nm}$, $\lambda_k = 90\,\mu\mathrm{m}$ and $\psi = 0^{\circ}$.}
    \label{Figure Reflection at Psi = 0}
\end{figure*}

	\subsection{Reflection}
    The wavevectors of the incident and reflected spin waves with frequency $\omega$ are denoted by $\mathbf{k} = (k_x,k_y) = k (\cos \phi, \sin \phi)$ and $\mathbf{k}' = (k_x',k_y') = k' (\cos \phi', \sin \phi')$, respectively. Due to the translational invariance of the system in the $y$-direction, we require that $k_y = k_y'$.  Now, to determine the relationship between the $x$-components of the wavevectors, we use the fact that the spin-wave frequency $\omega$ is required to be the same everywhere in the system. For a spin wave that is incident from the bare FMI, this relationship is given by
	\begin{align}\label{Equation Frequency Matching Reflection Insulator}
		\omega_0(\mathbf{k}) = \omega_0(\mathbf{k}'),
	\end{align}
    subject to the constraint $[\nabla_{\mathbf{k}} \omega_0(\mathbf{k})]_x\cdot[\nabla_{\mathbf{k}'} \omega_0(\mathbf{k}')]_x < 0$, while for a spin wave that is incident from the FMI-SC heterostructure, we have
	\begin{align}\label{Equation Reflection Hybrid Condition}
		\omega(\mathbf{k}) = \omega(\mathbf{k}'),
	\end{align}
    subject to the constraint $[\nabla_{\mathbf{k}} \omega(\mathbf{k})]_x\cdot[\nabla_{\mathbf{k}'} \omega(\mathbf{k}')]_x < 0$. We note that these constraints merely state that the component of the group velocity normal to the interface should have a different sign for incident and reflected wavevectors. All relevant information regarding reflection is contained in the solutions to the above equations, which we now proceed to study in detail.

    \subsubsection{Law of reflection for $\psi = 0$.}

     \begin{figure*}[t]
   \includegraphics[width= 0.9\linewidth]{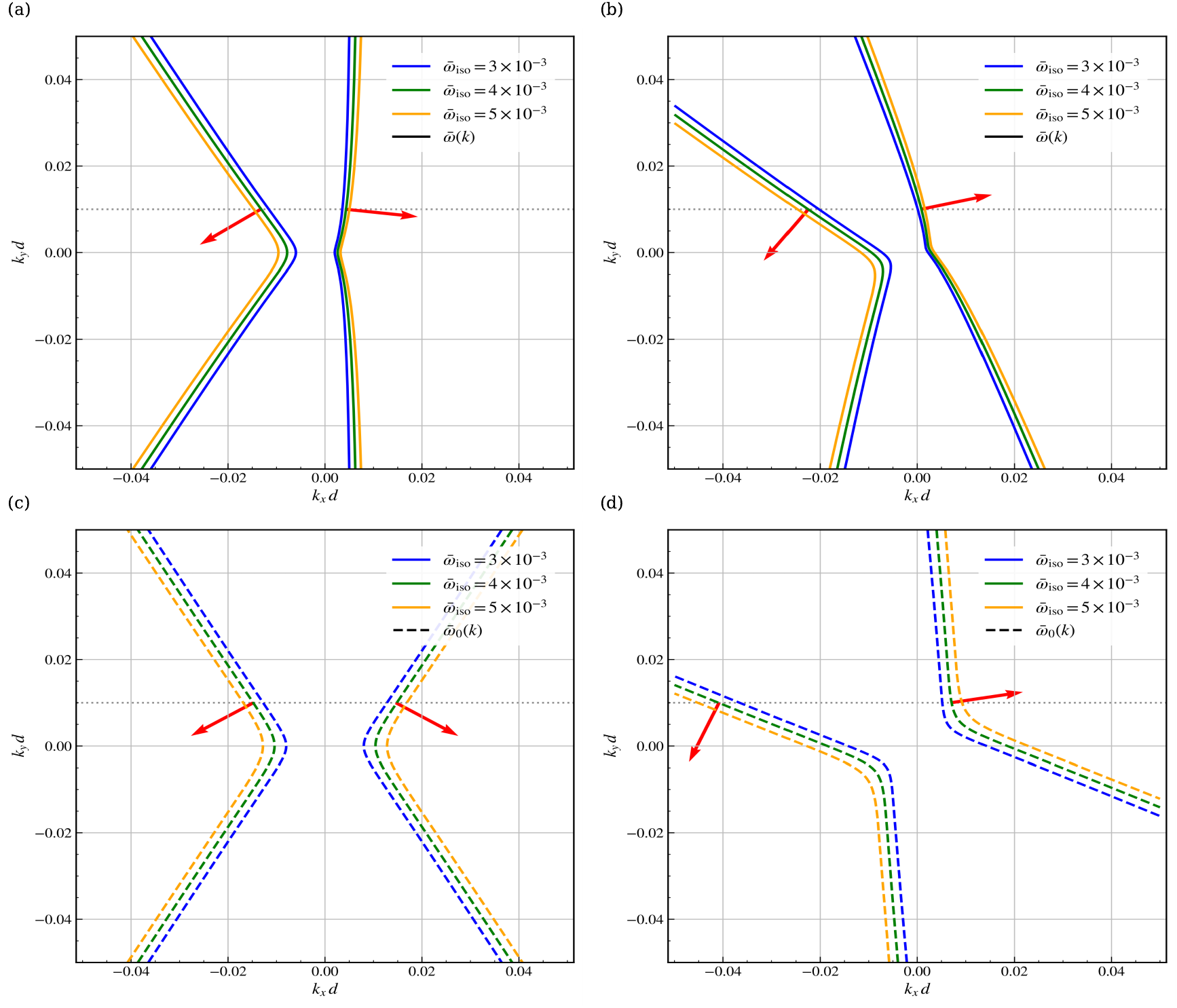}
    \caption{ \textbf{Negative reflection.} Plots of the isofrequency contours of the dispersion relations $\omega(\mathbf{k})$ and $\omega_0(\mathbf{k})$ in the London dipolar limit, given by Eqs.~\eqref{Equation Approximate Dispersion Hybrid} and \eqref{Equation Approximate Dispersion Bare Insulator}, respectively, for different values of the saturation magnetization angle $\psi$. The plots illustrate the emergence of negative reflection as $\psi$ is varied. For convenience, we have introduced the dimensionless frequencies $\bar{\omega}(\mathbf{k}) = [\omega(\mathbf{k})-\gamma B_0 \xi]/(\gamma \mu_0 M_S)$ and $\bar{\omega}_0(\mathbf{k}) = [\omega_0(\mathbf{k})-\gamma B_0 \xi]/(\gamma \mu_0 M_S)$, and in all panels we have set the ellipticity $\xi = 2$. The red arrows indicate the direction of the spin-wave group velocity at the points where the isofrequency contours intersect the gray dotted line $k_y d = 0.01$. Due to the translational invariance of the system in the $y$-direction, the incident and reflected spin waves have the same $k_y$. (a) Isofrequency contours of $\omega(\mathbf{k})$ for $\psi = 0^{\circ}$. Negative reflection does not occur, as the $y$-component of the group velocity has the same sign for the incident and reflected spin waves. (b) Isofrequency contours of $\omega(\mathbf{k})$ for $\psi = 20^{\circ}$. Negative reflection occurs, as both the $x$- and $y$-components of the group velocity have opposite signs for the incident and reflected spin waves. (c) Isofrequency contours of $\omega_0(\mathbf{k})$ for $\psi = 0^{\circ}$. The conventional law of reflection is obeyed, and no negative reflection occurs. (d) Isofrequency contours of $\omega_0(\mathbf{k})$ for $\psi = 35^{\circ}$. Negative reflection occurs, as both the $x$- and $y$-components of the group velocity have opposite signs for the incident and reflected spin waves. }
    \label{Figure Negative Reflection}
    \end{figure*}

	We first consider the case where $\psi = 0$, i.e., when the applied field $\mathbf{H}_0$ is parallel to the interface, for we can then capture the reflection laws by fairly simple formulas. 
    
    The reflection law in the bare FMI can be derived directly from either Eq. \eqref{Equation Dispersion Bare Insulator} or Eq.~\eqref{Equation Approximate Dispersion Bare Insulator}, and we have
	\begin{align}
		\phi' &= \pi-\phi,\\
		k' &= k.
	\end{align}
    We recognize this as the conventional law of reflection as encountered in classical optics. It should be stressed, however, that this law is valid only when $\psi = 0$, and highly nontrivial reflection behavior is observed in the situation where $\psi \neq 0$, as discussed in detail later. 

    For reflection in the FMI-SC heterostructure, unconventional properties are already present in the case where $\psi = 0$.
	We numerically solve Eq.~\eqref{Equation Reflection Hybrid Condition} using the exact dispersion relations for $\omega$, which in general yields one reflection solution. In the limit where $k d, k \lambda_L \ll 1$ and $d_s \gtrsim \lambda_L$, accurate results can also be obtained by using the approximated dispersion relation of Eq.~\eqref{Equation Approximate Dispersion Hybrid}. Plugging said result into Eq.~\eqref{Equation Reflection Hybrid Condition}, we can then find $k_x'$ by determining the real roots of the polynomial $f_{\mathbf{k}}(k_x')$, given by
	\begin{align}\label{Equation Reflection Polynomial}
		f_{\mathbf{k}}(k_x')&= Q_{f,\mathbf{k}}(k_x') + R_{f,\mathbf{k}}(k_x'),
	\end{align}
	where
	\begin{align}
		Q_{f,\mathbf{k}}(k_x') &= (\xi^2-1) k_x'^4 +2 C_{f,\mathbf{k}} k_x'^3-\left(C_{f,\mathbf{k}}^2+k_y^2\right)k_x'^2,\\
		R_{f,\mathbf{k}}(k_x') &=  C_{f,\mathbf{k}} k_y^2\left(2 k_x'-C_{f,\mathbf{k}} \right).
	\end{align}
	Here, we have defined the $\mathbf{k}$-dependent coefficient $C_{f,\mathbf{k}}$ as
	\begin{align}
		C_{f,\mathbf{k}} &= k_x\left( 1+\xi \dfrac{k_x}{k}\right),
	\end{align}
	and the physically relevant reflection solution is again selected by requiring that there is a sign difference between the $x$-components of the incident and outgoing group velocities, $\nabla_{\mathbf{k}'} \omega(\mathbf{k}')$ and $\nabla_{\mathbf{k}} \omega(\mathbf{k})$, respectively. 
	
	Now, the main advantage of this polynomial approximation is that it directly reveals to us that the ellipticity $\xi$ is the relevant parameter that determines the reflection in this regime. Furthermore, since $f_{\mathbf{k}}(k_x')$ is a homogeneous function of degree 4 in $k$ and $k_{x}'$, which means that $f_{\lambda\mathbf{k}}(\lambda k_x') = \lambda^4 f_{\mathbf{k}}(k_x')$ for any $\lambda \in \mathbb{C}$, it immediately follows that the roots of $f_{\mathbf{k}}(k_x')$ scale linearly with $k$, which in turn allows us to write down reflection laws that do not depend on the wavevector magnitude $k$. As indicated in Fig. \ref{Figure Reflection at Psi = 0}(a), a good agreement between the results derived from the exact and approximate dispersion relations is obtained, and we find that the reflection law for $|\phi| \leq 90^{\circ}$ (right-moving incident spin waves) is accurately described by the $k$-independent equation
	\begin{align}\label{Equation Reflection Law}
		\sin \phi' &\approx a_R(\xi) \sinh\bigg( b_R(\xi) \sin \phi \bigg),
	\end{align}
	where
	\begin{align}
		a_R(\xi) &\approx 0.07+0.57 \ln[0.55+0.23\,\xi]+0.013\,\xi, \\
		b_R(\xi) &\approx  \dfrac{3.16}{\sqrt{\xi-0.73}}.
	\end{align}
	We have numerically verified the validity of the reflection law for $\xi \geq 2$, or equivalently for $\mu_0 M_S \geq 3 B_0$. When the Zeeman interaction is significantly stronger than the dipolar interaction, $B_0 \gg \mu_0 M_S$, the aforementioned reflection law breaks down. 

    The most profound feature of Eq.~\eqref{Equation Reflection Law} is its universality, since there are no dependencies on material parameters such as the film thicknesses and the London penetration depth present anymore, the only relevant quantity being the ratio between the saturation magnetization and the magnitude of the static applied field. The coefficients in the aforementioned reflection law are therefore universal and, in the absence of intrinsic anisotropies, independent of the microscopic details of the system. The exact dependency of the reflection on $\xi$ is captured in the equations for $a_R(\xi)$ and $b_R(\xi)$, and illustrated in Fig. \ref{Figure Reflection at Psi = 0}(b).

     \begin{figure*}[t]
   \includegraphics[width= 0.9\linewidth]{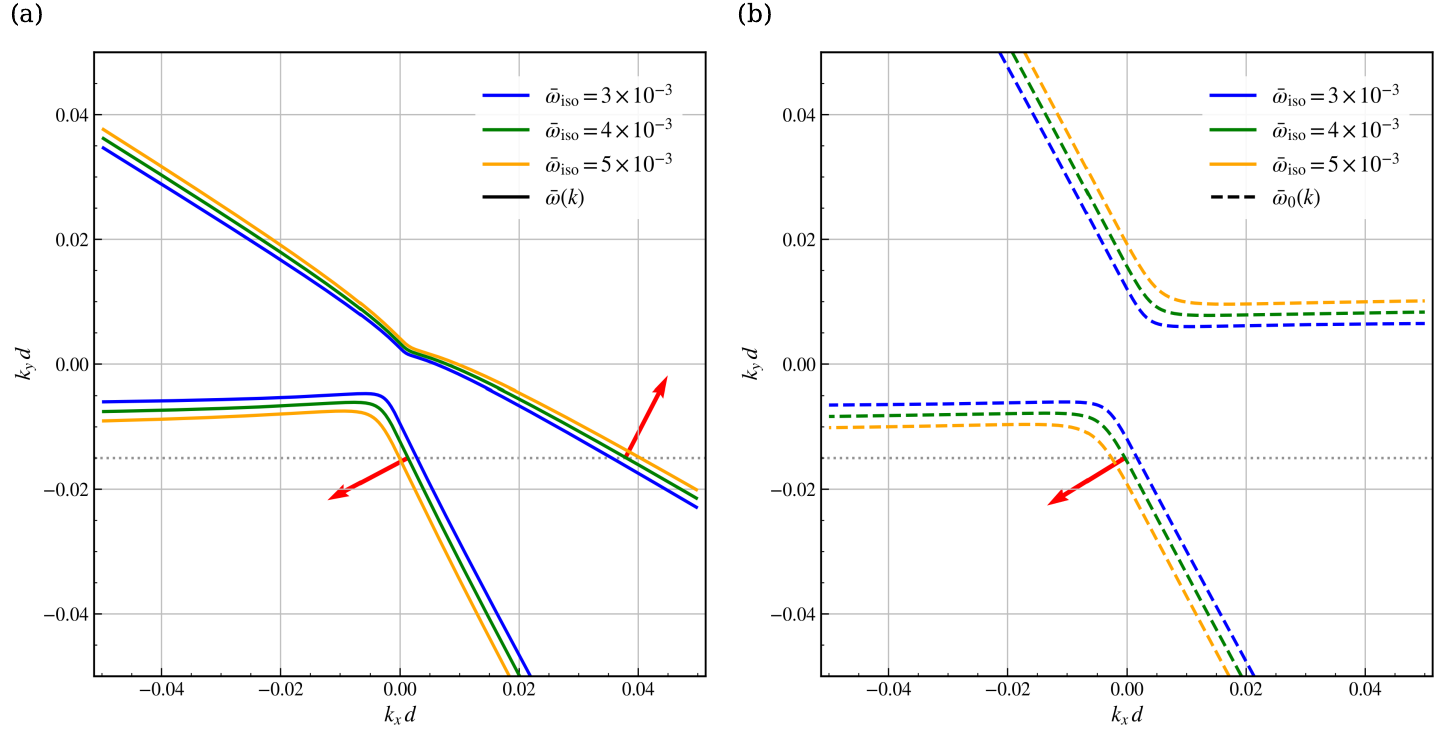}
    \caption{ \textbf{Temperature-controlled opening and closing of reflection channels}. Isofrequency contours of the dispersion relations $\omega(\mathbf{k})$ and $\omega_0(\mathbf{k})$ in the London dipolar limit, given by Eqs.~\eqref{Equation Approximate Dispersion Hybrid} and \eqref{Equation Approximate Dispersion Bare Insulator}, respectively, for ellipticity $\xi = 2$ and saturation magnetization angle $\psi = 60^{\circ}$. Here, $\omega(\mathbf{k})$ corresponds to the dispersion of the FMI-SC heterostructure in the superconducting state ($T<T_c$), while $\omega_0(\mathbf{k})$ corresponds to the dispersion of the same heterostructure in the normal state ($T>T_c$), where the Meissner screening is absent. For convenience, we have introduced the dimensionless frequencies $\bar{\omega}(\mathbf{k}) = [\omega(\mathbf{k})-\gamma B_0 \xi]/(\gamma \mu_0 M_S)$ and $\bar{\omega}_0(\mathbf{k}) = [\omega_0(\mathbf{k})-\gamma B_0 \xi]/(\gamma \mu_0 M_S)$. The red arrows indicate the direction of the spin-wave group velocity at the points where the isofrequency contours intersect the gray dotted line $k_y d = -0.015$. Due to the translational invariance of the system in the $y$-direction, the incident and reflected spin waves have the same $k_y$.(a) Isofrequency contours of $\omega(\mathbf{k})$. A (negative) reflection channel is available for spin waves with wavevector component $k_y d = -0.015$. (b) Isofrequency contours of $\omega_0(\mathbf{k})$. No reflection channels are available for spin waves with wavevector component $k_y d = -0.015$.}
    \label{Figure Reflection Channels}
    \end{figure*}
	
	The main consequence of Eq.~\eqref{Equation Reflection Law} is that the incoming and reflected angles are not related via a linear dependency, and that the specific value of $\phi'$ can be tuned by changing the magnitude $B_0$ of the static field. For example, when the incoming angle $\phi$ is around $50^{\circ}$, we find that $\phi' \approx 158^{\circ}$ for $\xi = 2$, while for $\xi = 4$ we find $\phi' \approx 144^{\circ}$. Furthermore, using the fact that the wavevector component along $y$ is conserved, we also find that the magnitude of $\mathbf{k}'$ is given by
	\begin{align}\label{Equation Wavevector Magnitude Reflection}
		k'(\phi) &\approx \dfrac{k \sin \phi}{a_R(\xi) \sinh\bigg( b_R(\xi) \sin \phi \bigg)},
	\end{align}
    and we again find that the relative size of $k'$ with respect to $k$ can be tuned significantly by changing $B_0$. In Fig. \ref{Figure Reflection at Psi = 0}(c), a plot of $k'/k$ as a function of the incident angle $\phi$ is shown for $\xi = 4$.

   Having discussed the reflection properties in the London dipolar limit, we now consider how the reflection in the FMI-SC system can be controlled by temperature. As mentioned previously, the London penetration depth depends on the temperature $T$ and is described by the Casimir-Gorter equation, Eq.~\eqref{Equation Casimir-Gorter}. A change in temperature therefore modifies the ratio $d_s/\lambda_L$, which in turn affects the spin-wave reflection in the FMI-SC system, as shown in Fig. \ref{Figure Reflection at Psi = 0}(d). As expected, when the London penetration depth is much larger than the thickness of the superconducting film, the conventional reflection law is recovered, whereas unconventional reflection properties emerge when the film thickness approaches or exceeds the London penetration depth. As illustrated in the inset of Fig. \ref{Figure Reflection at Psi = 0}(d), varying the temperature can change the reflected wavevector angle $\phi'$ by up to $30^{\circ}$, demonstrating the significant potential of the FMI-SC system for temperature-controlled spin-wave optics applications. Further discussions of these applications are provided in Section \ref{Section Applications}.

    Finally, before moving on to discuss anomalous reflection effects that occur when $\psi \neq 0$, an important comment needs to be made about the polynomial approach. As a consequence of the complex conjugate root theorem, $f_{\mathbf{k}}(k_x')$ can have either $0$, $2$ or $4$ real roots. In most cases, both the exact and approximate dispersion relations yield two real roots, which agree very well with each other. However, for large incident angles $\phi$, it can happen that the exact dispersion continues to produce two real roots, while the polynomial approach might introduce two additional spurious real roots alongside the correct ones. This behavior can be understood by noting that for large $\phi$, the coefficient $C_{f,\mathbf{k}}$ becomes very small. Consequently, truncating the dispersion relation to first order in $k d$ and $k \lambda_L$ is no longer adequate in this regime, as higher-order terms become non-negligible. In practice, the safest approach is thus to rely on the exact dispersion relation and solve the resulting transcendental equation numerically to determine the reflected wavevector $\mathbf{k}'$, while using the polynomial approximation mostly to gain qualitative insight into the parameters that influence the reflection in the limit where $k d, k \lambda_L \ll 1$ and $d_s \gtrsim \lambda_L$.

    \subsubsection{Negative reflection and reflection-channel control}

As briefly mentioned in the previous paragraph, the system exhibits even more intriguing properties when the angle $\psi$, which characterizes the orientation of the saturation magnetization, is nonzero.

The first phenomenon we discuss is \textit{negative reflection}. In optics, the conventional law of reflection dictates that the reflected light ray emerges on the opposite side of the interface normal compared to the incident ray. In contrast, negative reflection refers to the situation in which the reflected ray emerges on the same side of the interface normal as the incident ray. Negative reflection has been observed in optical metamaterials \cite{Zhang2007,Alvarez-Perez2022}, where it provides a mechanism for controlling the direction and spatial profile of reflected waves through engineered interfaces, as well as in spin-wave systems involving thin magnetic films with perpendicular magnetic anisotropy \cite{Lesniewski2026}, where it enables additional control over spin-wave propagation. In spin-wave systems, negative reflection is more precisely defined as the reflection of a wave whose normal and tangential components of the group velocity are directed opposite to the corresponding components of the incident spin wave.

A necessary, though not sufficient, condition for negative reflection is an anisotropic dispersion relation, as negative reflection cannot occur for an isotropic dispersion. Although the dispersions of both the bare FMI and the hybrid FMI-SC heterostructure considered here are strongly anisotropic, the corresponding isofrequency contours shown in Figs.~\ref{Figure Negative Reflection}(a) and (c) demonstrate that negative reflection does not occur for $\psi = 0$, as can also be shown directly from Eqs.~\eqref{Equation Approximate Dispersion Hybrid} and \eqref{Equation Approximate Dispersion Bare Insulator}. In addition to anisotropy, a second requirement is that the dispersion branches corresponding to the incident and reflected solutions must `bend towards each other'. Such bending can be induced by appropriately tuning the saturation magnetization angle $\psi$, resulting in negative reflection as illustrated in Figs.~\ref{Figure Negative Reflection}(b) and (d).

It should be emphasized that negative reflection can occur in both the bare FMI and the hybrid FMI-SC heterostructure, as is evident from the above mentioned figures. That being said, the Meissner screening fields of the superconductor typically lead to a stronger bending of one branch of the dispersion, thereby allowing negative reflection to occur for smaller values of $\psi$ than in the bare FMI.

The second phenomenon of interest is \textit{temperature-dependent reflection-channel control}. As discussed in the previous subsection, the reflection angle depends sensitively on the ratio between the superconducting film thickness and the London penetration depth, and can therefore be tuned by temperature. Here, we demonstrate an even more striking effect, namely that, by an appropriate choice of the saturation magnetization angle $\psi$, temperature can be used not only to modify the reflection angle, but also to open or close reflection channels altogether. This is illustrated in Fig. \ref{Figure Reflection Channels}, in which the isofrequency contours of $\omega(\mathbf{k})$ and $\omega_0(\mathbf{k})$ for $\psi = 60^{\circ}$ and $\xi = 2$ are plotted. Here, we note that $\omega_0(\mathbf{k})$ can also be interpreted as the dispersion of the FMI-SC heterostructure above the superconducting critical temperature, $T>T_c$. As can be seen from the figure, for the given $y$-component of the spin-wave momentum, $k_y$, a (negative) reflection channel exists for $T<T_c$, whereas this channel is closed for $T>T_c$ \footnote{Strictly speaking, a reflection solution may still exist, as the exchange contribution to the frequency causes the isofrequency contours to close eventually. This solution, however, will be located at $k \gg 1/d$, and therefore lies far outside the dipolar regime.}. This ability to thermally control both the reflection angle and the existence of reflection channels opens possibilities for tunable spin-wave mirrors, which could have potential applications for spin-wave interferometers, as discussed in more detail in Section \ref{Section Applications}.

	\subsection{Refraction}
    We now continue with a discussion on refraction properties. The wavevectors of the incident and refracted spin waves with frequency $\omega$ are denoted by $\mathbf{k} = (k_x,k_y) = k (\cos \phi, \sin \phi)$ and $\mathbf{k}' = (k_x',k_y') = k' (\cos \phi', \sin \phi')$, respectively.
	Proceeding in a similar manner as we did when studying reflection, we can determine $\mathbf{k}'$ by solving the equation
	\begin{align}\label{Equation Frequency Matching Refraction Insulator}
		\omega(\mathbf{k}') = \omega_0(\mathbf{k}),
	\end{align}
	subject to the constraint $k_y' = k_y$. Here, we explicitly assume that the incident spin wave comes from the bare MI, while the refracted wave is located in the FMI-SC heterostructure. The physically relevant refraction solution is selected by requiring that the $x$-components of the incident and outgoing group velocities, $\nabla_{\mathbf{k}'} \omega(\mathbf{k}')$ and $\nabla_{\mathbf{k}} \omega_0(\mathbf{k})$, have the same sign. All relevant information regarding refraction is contained in the solutions to the above equations. In what follows, we focus exclusively on right-propagating spin waves, i.e., spin waves with a positive $x$-component of the group velocity, for reasons that will become clear later.

    \subsubsection{Law of refraction for $\psi = 0$}

    \begin{figure*}[t]
   \includegraphics[width=0.9\linewidth]{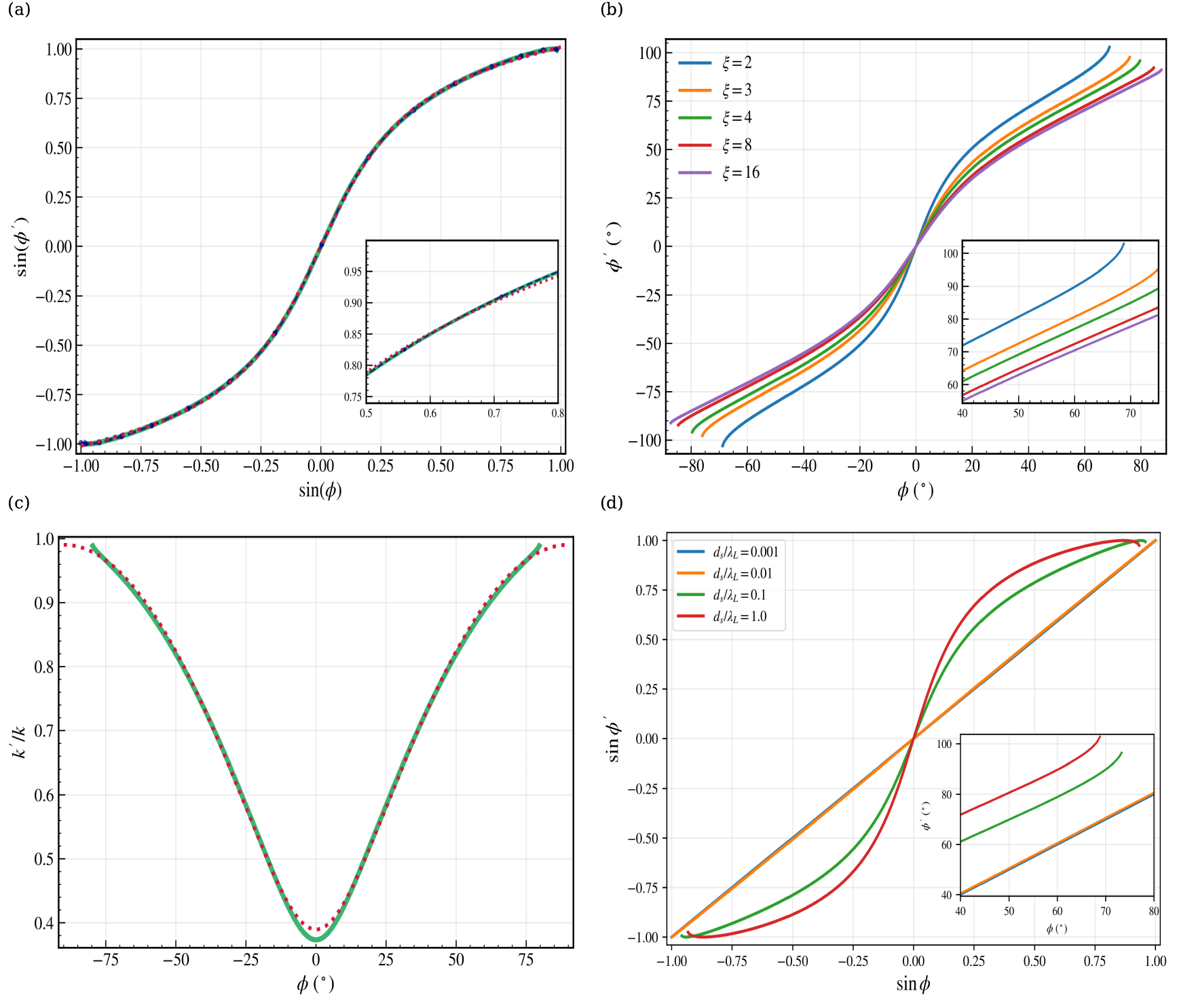}
    \caption{\textbf{Refraction law and negative phase refraction.} (a) Sine of the angle $\phi'$ of the refracted wavevector as a function of the sine of the angle $\phi$ of the incident wavevector. The green, solid line shows the result obtained from the exact dispersion relations, Eqs.~\eqref{Equation Spin Wave Dispersion} and \eqref{Equation Dispersion Bare Insulator}, while the dark blue, dashed line is obtained using the polynomial approximations, Eqs.~\eqref{Equation Reflection Polynomial} and \eqref{Equation Refraction Polynomial}. The red, dotted line represents the approximate refraction law given in Eq.~\eqref{Equation Refraction Law}. (b) Refracted angle $\phi'$ as a function of the incident angle $\phi$ for different values of the ellipticity $\xi$. The inset shows the occurrence of negative phase refraction. (c) Ratio between the magnitudes of the refracted ($k'$) and incident ($k$) wavevectors as a function of the incident angle $\phi$. The green, solid line shows the result obtained from the exact dispersion relations, Eqs.~\eqref{Equation Spin Wave Dispersion} and \eqref{Equation Dispersion Bare Insulator}, while the red, dotted line indicates the result given by Eq.~\eqref{Equation Wavevector Magnitude Refraction}. (d) Sine of the angle $\phi'$ of the refracted wavevector as a function of the sine of the angle $\phi$ of the incident wavevector for different values of $d_s/\lambda_L$, with $d_s = 100$ nm and $\xi = 2$. In all plots, incident spin waves have wavevector magnitude $k = 2\pi/\lambda_k$ and a group velocity with a positive $x$-component. The spin waves are assumed to be incident from the bare FMI side. Unless stated otherwise, the following parameters were used: $D=0.053\,\mathrm{mT}\,\mu \mathrm{m}^2$, $\mu_0 M_S = 250\,\mathrm{mT}$, $\xi = 4$, $d = 100\,\mathrm{nm}$, $d_s = 1\,\mu\mathrm{m}$, $\lambda_L = 100\,\mathrm{nm}$, $\lambda_k = 90\,\mu\mathrm{m}$, and $\psi = 0^{\circ}$.}
    \label{Figure Refraction at Psi = 0}
\end{figure*}

    Like before, setting $\psi = 0$ and using the approximated dispersion relations of Eq.~\eqref{Equation Approximate Dispersion Hybrid} and \eqref{Equation Approximate Dispersion Bare Insulator}, we can gain useful insight into the refraction by considering the roots of the polynomial $g_{\mathbf{k}}(k_x')$, given by
	\begin{align}\label{Equation Refraction Polynomial}
		g_{\mathbf{k}}(k_x')&= Q_{g,\mathbf{k}}(k_x') + R_{g,\mathbf{k}}(k_x'),
	\end{align}
	where
	\begin{align}
		Q_{g,\mathbf{k}}(k_x') &= (\xi^2-1) k_x'^4 +2 C_{g,\mathbf{k}} k_x'^3-\left(C_{g,\mathbf{k}}^2+k_y^2\right)k_x'^2,\\
		R_{g,\mathbf{k}}(k_x') &=  C_{g,\mathbf{k}} k_y^2\left(2 k_x'-C_{g,\mathbf{k}} \right).
	\end{align}
	Here, we have defined the $\mathbf{k}$-dependent coefficient $C_{g,\mathbf{k}}$ as
	\begin{align}
		C_{g,\mathbf{k}} &= \frac{k}{2\xi}\left( -1+\xi^2 \dfrac{k_x^2}{k^2}\right).
	\end{align}
    As indicated in Fig. \ref{Figure Refraction at Psi = 0}(a), we again obtain a good agreement between the results derived from the exact and approximate dispersion relations. From the homogeneity of $g_{\mathbf{k}}(k_x')$, we then arrive at the following $k$-independent refraction law, 
    \begin{align}\label{Equation Refraction Law}
		\sin \phi' &\approx \dfrac{a_T(\xi) \sin \phi}{[1+b_T(\xi)\sin^2(\phi)]^{p(\xi)}} 
	\end{align}
	where
	\begin{align}
		a_T(\xi) &\approx 2.01+\dfrac{1.70}{\xi-0.95} , \\
		b_T(\xi) &\approx 13.0+120\,e^{-1.70\xi}-4.21\,\xi^{-0.18},\\
        p(\xi) &\approx 0.28+\dfrac{0.47}{\xi+0.20}.
	\end{align}
    We have numerically verified the validity of this law for $\xi \geq 2$, or equivalently for $\mu_0 M_S \geq 3 B_0$. We note that the refraction law reduces to the familiar Snell's law form, 
    \begin{align}
		\sin \phi' = \dfrac{n(\xi)}{n'(\xi)} \sin \phi,
	\end{align}
    for $|\phi| \ll 1$, with the refractive index ratio $n(\xi)/n'(\xi)$ given by
    \begin{align}
        \dfrac{n(\xi)}{n'(\xi)} &= a_T(\xi).
    \end{align}
    For large incident angles, however, strong deviations from Snell's law are observed. As was the case for spin-wave reflection, we find that the angle of refraction, $\phi'$, strongly depends on the ratio between the static field amplitude $B_0$ and the saturation magnetization $M_S$. For example, when the incident angle is given by $\phi = 50^{\circ}$, we find that $\phi' \approx 81^{\circ}$  for $\xi = 2$, while for $\xi = 4$ we obtain $\phi' \approx 69^{\circ}$.

    Perhaps even more interesting is the occurrence of \textit{negative phase refraction}, illustrated most clearly in the inset of Fig.~\ref{Figure Refraction at Psi = 0}(b). There, it is shown that the refracted wavevector $\mathbf{k}'$ can emerge with an angle $\phi' > 90^{\circ}$, even though the incident wavevector $\mathbf{k}$ satisfies $|\phi| < 90^{\circ}$. Consequently, both $\mathbf{k}$ and $\mathbf{k}'$ point toward the interface, which implies that the phase velocities of both waves are directed inward, as the wavevector is normal to surfaces of constant phase.

    This phenomenon, known as negative phase refraction, was first analyzed in detail by Veselago in his seminal work on electromagnetic wave propagation in media with simultaneously negative permittivity and permeability \cite{Veselago1968}. It is closely related to, but distinct from, \textit{negative refraction} of the group velocity (or energy flow), which we discuss in the next subsection. This distinction is important because the phase velocity and the group velocity need not point in the same direction. In optics, negative phase and group velocity refraction have inspired a wide range of applications, including Pendry's proposal of a superlens \cite{Pendry2000}, capable of subwavelength imaging, as well as concepts for transformation-optics devices such as invisibility cloaks and other unconventional wave-guiding structures \cite{Pendry2006,Leonhardt2006,Rahm2008,Chen2010}.

    Having discussed negative phase refraction, we emphasize that, just like the reflection law, the law of refraction also obeys a certain universality principle in the London dipolar limit. In this limit, Eq.~\eqref{Equation Refraction Law} shows that the refraction depends only on the ellipticity $\xi$, making it independent of microscopic material parameters such as the film thicknesses and the London penetration depth. The corresponding dependence on $\xi$ is shown in Fig.~\ref{Figure Refraction at Psi = 0}(b).

\begin{figure}
        \centering
        \includegraphics[width=1\linewidth]{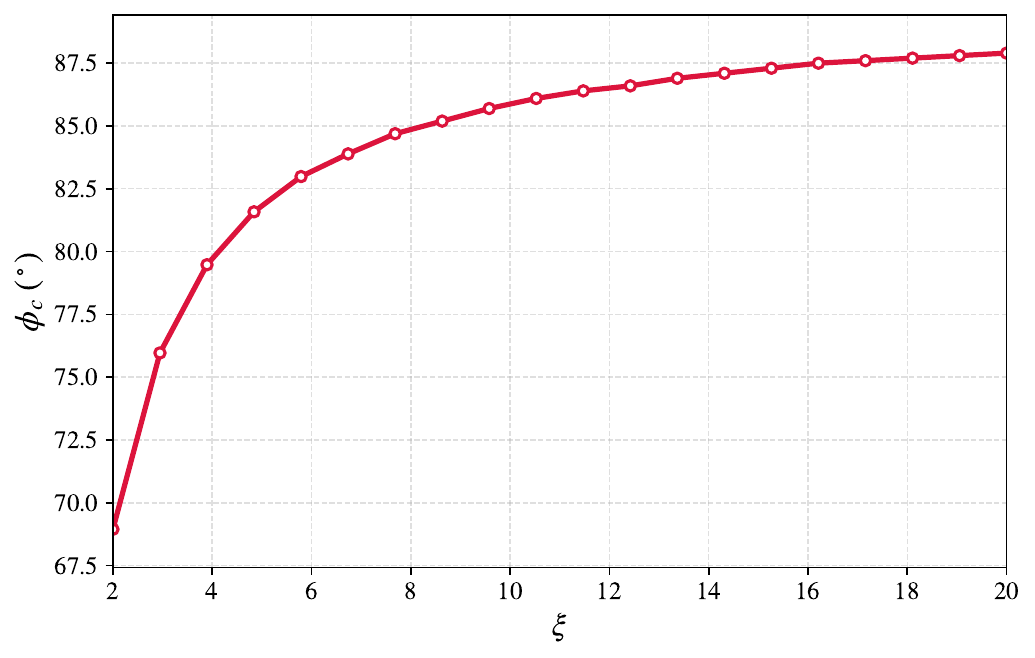}
        \caption{\textbf{Critical refraction angles}. The critical angle of refraction, $\phi_c$, in the London dipolar limit is plotted as a function of the ellipticity $\xi$. The critical angle increases monotonically with $\xi$, approaching a limiting value of $\sim 88^\circ$ for large $\xi$. The following parameters were used to create this plot: $D=0.053\,\mathrm{mT}\,\mu \mathrm{m}^2$, $\mu_0 M_S = 250\,\mathrm{mT}$, $\xi = 4$, $d = 100\,\mathrm{nm}$, $d_s = 1\,\mu\mathrm{m}$, $\lambda_L = 100\,\mathrm{nm}$, $\lambda_k = 90\,\mu\mathrm{m}$, and $\psi = 0^{\circ}$.  }
        \label{Figure Critical Angle}
    \end{figure}

    It should be stressed, however, that Eq.~\eqref{Equation Refraction Law} is only valid for $|\phi| < \phi_c$, where $\phi_c$ is the \textit{critical angle}. The critical angle is defined as the largest (positive) angle of incidence for which a refracted wave solution exist. Similar to the discussion of the reflection law, this critical angle cannot be obtained within the polynomial approximation and instead requires solving the exact transcendental dispersion relation numerically. The breakdown of the polynomial approach is again associated with the increasing importance of higher-order terms in the dispersion relation at large incident angles. Within the London dipolar limit regime, where $k d, k \lambda_L \ll 1$ and $d_s \gtrsim \lambda_L$, the critical angle still depends predominantly on the ellipticity $\xi$, as illustrated in Fig.~\ref{Figure Critical Angle}.

    A similar law, describing the dependence of the magnitude of the refracted wavevector $\mathbf{k}'$ on the incident angle $\phi$, can be obtained directly, and we find that
    \begin{equation}\label{Equation Wavevector Magnitude Refraction}
        k'(\phi) \approx \dfrac{k \, [1+b_T(\xi)\sin^2(\phi)]^{p(\xi)}}{a_T(\xi)}.
    \end{equation}
    In Fig. \ref{Figure Refraction at Psi = 0}(c), a plot of $k'/k$ is shown as a function of $\phi$ for $\xi = 4$. Similar to experimental observations in Ref.~\cite{Borst2023}, we see that the wavelength of spin waves at a given frequency $\omega$ is typically larger in the FMI-SC heterostructure than in the bare MI, in particular at normal incidence.

     \begin{figure*}[t]
   \includegraphics[width= 0.9\linewidth]{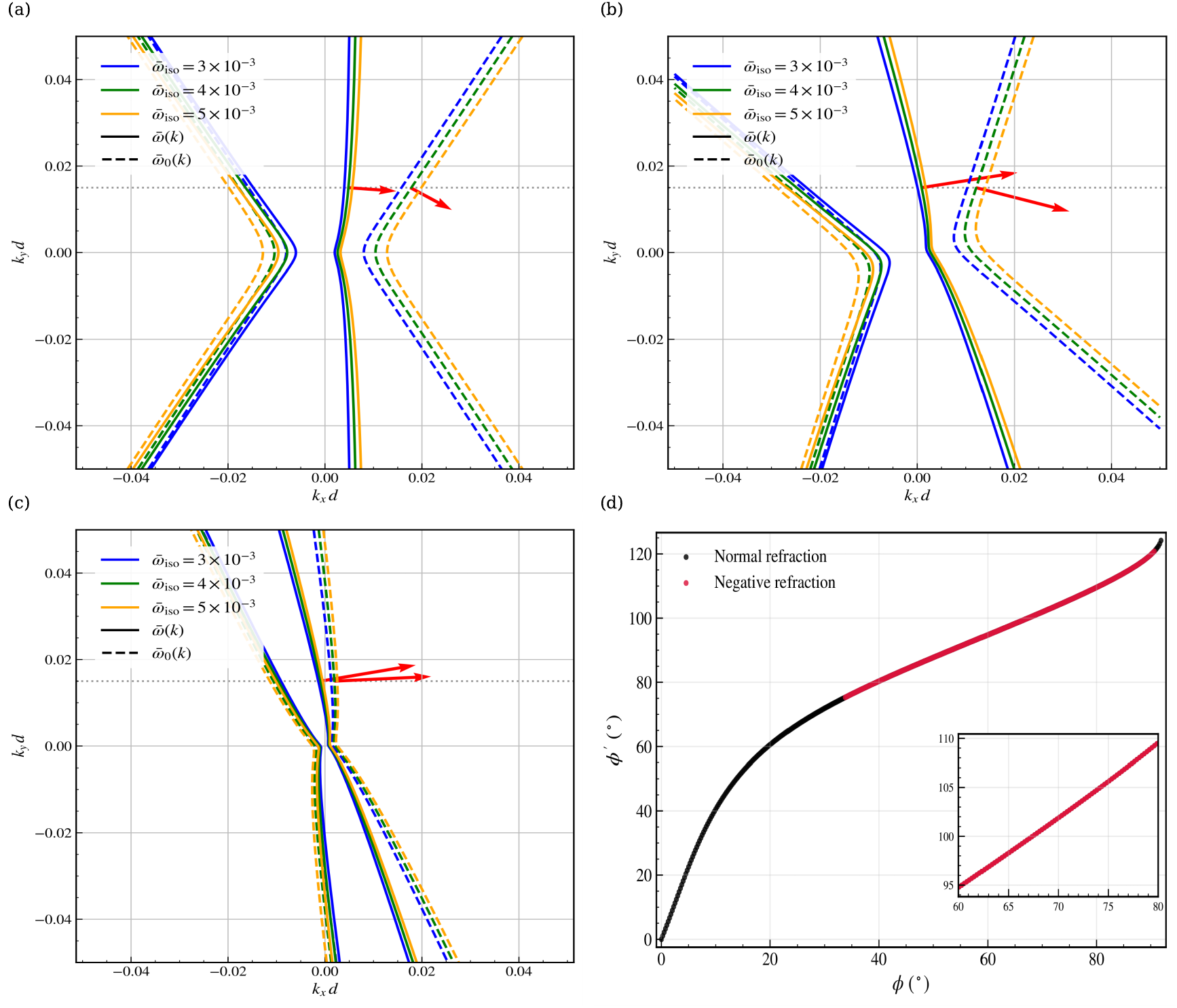}
    \caption{ \textbf{Negative refraction.} Plots of the isofrequency contours of the dispersion relations $\omega(\mathbf{k})$ and $\omega_0(\mathbf{k})$ in the London dipolar limit, given by Eqs.~\eqref{Equation Approximate Dispersion Hybrid} and \eqref{Equation Approximate Dispersion Bare Insulator}, respectively, together with a refraction diagram in a typical negative-refraction regime. The plots illustrate the emergence of negative refraction with respect to the group velocity for right-moving spin waves as the saturation magnetization angle $\psi$ and the ellipticity $\xi$ are varied. For convenience, we have introduced the dimensionless frequencies $\bar{\omega}(\mathbf{k}) = [\omega(\mathbf{k})-\gamma B_0 \xi]/(\gamma \mu_0 M_S)$ and $\bar{\omega}_0(\mathbf{k}) = [\omega_0(\mathbf{k})-\gamma B_0 \xi]/(\gamma \mu_0 M_S)$. The red arrows indicate the direction of the spin-wave group velocity at the points where the rightmost isofrequency contours, corresponding to right-propagating spin waves, intersect the gray dotted line $k_y d = 0.015$. Due to the translational invariance of the system in the $y$-direction, the incident and refracted spin waves have the same $k_y$. (a) Isofrequency contours of $\omega(\mathbf{k})$ and $\omega_0(\mathbf{k})$ for $\psi = 0^{\circ}$ and $\xi = 2$. Negative refraction does not occur, as the $y$-component of the group velocity has the same sign for the incident and refracted spin waves. (b) Isofrequency contours of $\omega(\mathbf{k})$ and $\omega_0(\mathbf{k})$ for $\psi = 15^{\circ}$ and $\xi = 2$. Negative refraction occurs, since the $y$-component of the group velocity has opposite sign for the incident and refracted spin waves. (c) Isofrequency contours of $\omega(\mathbf{k})$ and $\omega_0(\mathbf{k})$ for $\psi = 15^{\circ}$ and $\xi = 20$. Negative refraction is suppressed as a result of the increase in $\xi$. (d) Refraction diagram showing the angle $\phi'$ that the wavevector $\mathbf{k}'$ of the refracted spin wave makes with the interface normal ($\hat{\mathbf{x}}$) as a function of the incident angle $\phi$, for an incident spin wave with wavevector magnitude $k = 10^{-2}/d$. As shown in the inset, negative refraction (with respect to the group velocity) and negative phase refraction typically occur simultaneously. Here, we used the exact dispersion relations, Eqs.~\eqref{Equation Spin Wave Dispersion} and \eqref{Equation Dispersion Bare Insulator}, with parameters $D=0.053\,\mathrm{mT\,\mu m^2}$, $\mu_0 M_S = 250\,\mathrm{mT}$, $\xi = 2$, and $d = d_s = \lambda_L = 100\,\mathrm{nm}$.   }
    \label{Figure Negative Refraction}
    \end{figure*}

    Finally, before discussing the anomalous refraction phenomena that emerge when $\psi \neq 0$, we consider the temperature-controlled tuning of refraction in the system. Similar to the case of reflection, variations in the temperature $T$ modify the ratio $d_s/\lambda_L$, thereby strongly influencing the refraction behavior, as shown in Fig. \ref{Figure Refraction at Psi = 0}(d). In the limit where the London penetration depth greatly exceeds the superconducting film thickness, we find that the incident and refracted angle are equal to one another, meaning negligible deflection takes place. However, as the film thickness becomes comparable to or larger than the London penetration depth, the unconventional refraction effects previously discussed gradually emerge. As shown in the inset of Fig. \ref{Figure Refraction at Psi = 0}(d), temperature variation enables tuning of the refracted wavevector angle $\phi'$ by up to $30^{\circ}$, again highlighting the considerable potential of the FMI-SC platform for temperature-controlled spin-wave optics.

    \subsubsection{Negative refraction}\label{Section Negative Refraction}

   We now consider refraction properties that emerge when the saturation magnetization angle $\psi$ is nonzero.

   The most intriguing phenomenon that occurs exclusively when $\psi \neq 0$ is \textit{negative refraction} with respect to the group velocity. Conventional optics dictates that the refracted light ray emerges on the opposite side of the interface normal compared to the incident ray. In contrast, negative refraction refers to the situation in which the refracted ray emerges on the same side of the interface normal as the incident ray. As mentioned previously, negative refraction has found a myriad of applications in transformation optics \cite{Pendry2000,Pendry2006,Leonhardt2006,Rahm2008,Chen2010}, including superlenses and optical cloaking, and is closely associated with unconventional phenomena such as the reversed Doppler effect and reversed Cherenkov radiation \cite{Veselago1968}. Negative refraction can also occur in spin-wave systems \cite{Kim2008,Hioki2020,Lesniewski2026}, where it is more precisely defined as the refraction of a wave whose tangential component of the group velocity is opposite in direction to that of the incident spin wave.

   Unlike negative reflection, the phenomenon of negative refraction can also occur for isotropic dispersion relations. For example, consider the refraction of a (spin) wave at the interface between two materials, denoted by $(+)$ and $(-)$, with respective dispersion relations
   \begin{align*}
       \omega_+(k) &= A_+ + B_+ \,k,\\
       \omega_-(k) &= A_- - B_- \,k,
   \end{align*}
   where $A_{\pm},B_{\pm} > 0$. A moment's thought shows that this system exhibits negative refraction. 
   
   Recalling that, in the London dipolar limit, the anisotropic spin-wave dispersions in our system are given by
   \begin{align*}
       \omega(\mathbf{k}) &\approx \gamma B_0 \xi + \dfrac{\gamma \mu_0 M_S}{2} k d \cos(\phi-\psi)\bigg[1+ \xi \cos(\phi-\psi) \bigg],\\
       \omega_0(\mathbf{k}) &\approx \gamma B_0 \xi + \dfrac{\gamma \mu_0 M_S}{4 \xi} k d \bigg[-1+ \xi^2 \cos^2(\phi-\psi) \bigg],
   \end{align*}
   it seems reasonable to expect that, due to the presence of the $\pm 1$ term inside the square brackets, negative refraction occurs here as well. Although negative refraction does indeed occur in this system, the anisotropy of the dispersion relation makes it far from obvious under what conditions it arises exactly. A calculation of the gradients of $\omega(\mathbf{k})$ and $\omega_0(\mathbf{k})$ shows that a necessary condition is $\psi \neq 0$, as illustrated in Figs.~\ref{Figure Negative Refraction}(a) and (b). Moreover, inspection of the dispersion relations indicates that the ellipticity $\xi$ cannot be too large, as otherwise the terms proportional to $\cos^2(\phi-\psi)$ dominate the dispersion and suppress negative refraction, as shown in Fig.~\ref{Figure Negative Refraction}(c).

   Negative refraction in our system is most clearly demonstrated in Fig.~\ref{Figure Negative Refraction}(b). As shown there, an appropriate choice of $\psi$ (here, $\psi = 15^{\circ}$) causes the dispersion branches corresponding to right-propagating spin waves to bend away from one another, thereby creating the conditions necessary for negative refraction to take place.
   From Figs.~\ref{Figure Negative Refraction}(a)–(c), it is also clear that negative refraction does not occur for left-propagating spin waves when $|\psi|<90^{\circ}$, since the dispersion branch $\omega(\mathbf{k})$ associated with left-propagating spin waves is only weakly affected by Meissner screening. This directional asymmetry can be understood from the nonreciprocal nature of dipolar spin waves, as the dynamic stray fields associated with opposite propagation directions are predominantly localized at opposite surfaces of the film. Because the superconductor is present only on one side of the magnetic film, its screening response is therefore stronger for one propagation direction than for the other, leading to a direction-dependent modification of the spin-wave dispersion. For $|\psi| > 90^{\circ}$, the roles are reversed, with left-propagating spin waves exhibiting negative refraction, while right-propagating spin waves remain largely unaffected.

   As indicated in Fig.~\ref{Figure Negative Refraction}(d), for an incident spin wave with wavevector $\mathbf{k} = k \,(\cos \phi, \sin \phi)$, negative refraction is not restricted to a few isolated incident angles $\phi$, but can occur over a broad range of angles. The extent of this range can be controlled by varying the parameters $\xi$ and $\psi$, thereby providing a flexible means of tuning the refraction behavior.

   Before moving on to the spin-wave Fresnel equations of the system, we briefly comment more broadly on the impact of Meissner screening on the spin-wave bands. Besides the fact that only one of the branches of $\omega(\mathbf{k})$ is strongly modified, we observe in Fig.~\ref{Figure Negative Refraction} that this affected branch becomes highly flattened along a particular direction in momentum space. More specifically, for values of $\xi$ sufficiently close to unity, the corresponding isofrequency contours become nearly straight lines oriented approximately parallel to
   \begin{align}
       k_y \sin \psi = \dfrac{2[\omega - \gamma B_0 \xi]}{\gamma \mu_0 M_Sd}- k_x \cos \psi.
   \end{align}
   In the language of electronic band engineering, this would correspond to a highly anisotropic effective mass tensor \cite{Ashcroft1976}. As a consequence, the direction of the group velocity, which can be tuned by $\psi$, becomes nearly independent of the incident wavevector, leading to a pronounced self-collimation of spin-wave propagation over a broad range of wavevectors. This self-collimation is a key ingredient for low-diffraction spin-wave transport and perfect imaging, as discussed in more detail in Section \ref{Section Applications}.

	\begin{figure*}[t]
        \centering
        \includegraphics[width=0.9\linewidth]{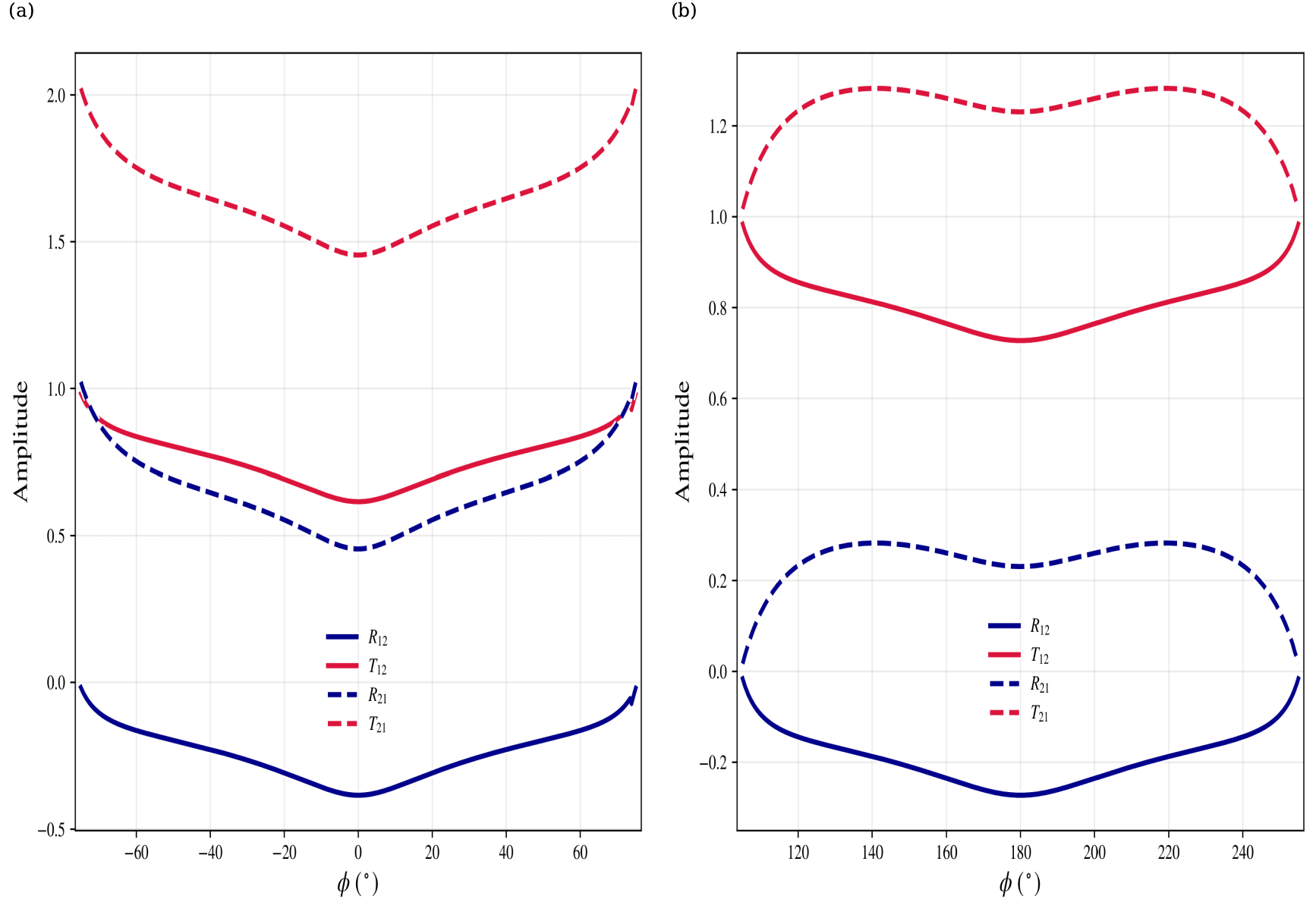}
        \caption{\textbf{Fresnel amplitudes in the London dipolar limit}. Plots of the Fresnel amplitudes $R_{12}$, $T_{12}$, $R_{21}$ and $T_{21}$, as given in Eqs.~\ref{Equation First Fresnel Equation}-\ref{Equation Last Fresnel Equation}, for $\psi = 0^{\circ}$ as a function of the incident angle $\phi$. (a) Fresnel amplitudes for incident spin waves with a positive $x$-component of the group velocity. (b) Fresnel amplitudes for incident spin waves with a negative $x$-component of the group velocity. In all plots, incident spin waves have wavevector magnitude $k = 2\pi/\lambda_k$, and we used the exact dispersion relations, Eqs.~\eqref{Equation Spin Wave Dispersion} and \eqref{Equation Dispersion Bare Insulator}, with parameters $D=0.053\,\mathrm{mT\,\mu m^2}$, $\mu_0 M_S = 250\,\mathrm{mT}$, $\xi = 4$, $d = d_s = \lambda_L = 100\,\mathrm{nm}$ and $\lambda_k = 90\,\mu\mathrm{m}$.}
        \label{Figure Fresnel Amplitudes}
    \end{figure*}
	
	\subsection{Spin-wave Fresnel equations}\label{Section Fresnel Equations}
    We now focus our attention on the reflection and transmission coefficients of the system. For the sake of algebraic simplicity, we focus exclusively on the case where the saturation magnetization angle $\psi$ is zero, i.e., when $\mathbf{B}_0 = B_0 \hat{\mathbf{y}}$. We consider two possible scattering configurations for the spin waves.
    In the first case, corresponding to a spin-wave magnetization density $\mathbf{m}_{12}(\mathbf{r}) e^{-i\omega t}$, the incident and reflected waves propagate in the bare FMI, while the transmitted wave propagates in the FMI-SC heterostructure, with the interface located at $x=0$. For $-d/2 \leq z \leq d/2$, we then have
    \begin{align}\label{Equation Scattering Magnetization 12}
        \mathbf{m}_{12}(\mathbf{r}) &=
        \begin{cases}
        e^{i\mathbf{k}\cdot\boldsymbol{\rho}}\,\mathbf{m}_0(\mathbf{k}) 
        + R_{12}
        e^{i\mathbf{k}_{ R}\cdot\boldsymbol{\rho}}\,
        \mathbf{m}_0(\mathbf{k}_{ R}) \hspace{0.2cm} (x < 0),\\ T_{12}
        e^{i\mathbf{k}_{T}\cdot\boldsymbol{\rho}}\,\mathbf{m}(\mathbf{k}_{ T}) \hspace{2.6cm} (x > 0),
        \end{cases}
    \end{align}
    where $\mathbf{k}$, $\mathbf{k}_R$ and $\mathbf{k}_T$ denote the wavevectors of the incident, reflected and transmitted spin waves, respectively, and where $R_{12} \in \mathbb{C}$ and $T_{12} \in \mathbb{C}$ correspond to the reflection and transmission amplitudes. The $\mathbf{k}$-dependent polarization vectors $\mathbf{m}_0(\mathbf{k})$ and $\mathbf{m}(\mathbf{k})$ are given by
    \begin{align}
        \mathbf{m}_0(\mathbf{k}) &=
        \begin{pmatrix}
            i \eta_0(\mathbf{k}) \\ 0 \\ 1
        \end{pmatrix}
        ,
    \end{align}
    and
    \begin{align}
        \mathbf{m}(\mathbf{k}) &=
        \begin{pmatrix}
            i \eta(\mathbf{k}) \\ 0 \\ 1
        \end{pmatrix}
        ,
    \end{align}
    respectively. In the second case, the propagation direction is reversed, such that the incident and reflected waves are located in the heterostructure and the transmitted wave propagates in the bare FMI. The spin-wave magnetization density is then given by $\mathbf{m}_{21}(\mathbf{r},t) e^{-i\omega t}$, with
    \begin{align}
        \mathbf{m}_{21}(\mathbf{r}) &=
        \begin{cases}
        T_{21}
        e^{i\mathbf{k}_{T}\cdot\boldsymbol{\rho}}\,\mathbf{m}_0(\mathbf{k}_{ T}) \hspace{2.2cm} (x < 0),\\ e^{i\mathbf{k}\cdot\boldsymbol{\rho}}\,\mathbf{m}(\mathbf{k}) 
        + R_{21}
        e^{i\mathbf{k}_{ R}\cdot\boldsymbol{\rho}}\,
        \mathbf{m}(\mathbf{k}_{ R}) \hspace{0.2cm} (x > 0),
        \end{cases}
    \end{align}
    for $-d/2 \leq z \leq d/2$. 

    Before solving the scattering problem, we note that the equations above have been written under the assumption that the incident spin waves originate from the left ($x<0$) in the FMI and from the right ($x>0$) in the FMI-SC. One might object that, owing to the strong anisotropy of the dispersion relations, the opposite configurations---namely, incident waves originating from the right (left) in the FMI (FMI-SC)---should be treated separately. This is not necessary, however, because we leave the reflected and transmitted wavevectors implicit, and the functional forms of the reflection and transmission coefficients are therefore identical for both propagation directions. Consequently, the coefficients $R_{12}$ and $T_{12}$ correspond to spin waves incident from the FMI, while $R_{21}$ and $T_{21}$ correspond to spin waves incident from the FMI-SC.

     To determine $R_{12}$ ($R_{21}$) and $T_{12}$ ($T_{21}$), we need to apply suitable boundary conditions to the spin-wave scattering problem. Since we work exclusively in the dipolar regime, we follow Ref.~\cite{Verba2020} and impose magnetostatic boundary conditions. Denoting the total magnetic field and the total magnetizing field inside the film by $\mathbf{B}^{\mathrm{in}}$ and $\mathbf{H}^{\mathrm{in}}$, respectively the boundary conditions are given by
     \begin{align}
         \lim_{x \rightarrow 0^-} \mathbf{B}^{\mathrm{in}}(\mathbf{r},t) \cdot \hat{\mathbf{x}} &= \lim_{x \rightarrow 0^+} \mathbf{B}^{\mathrm{in}}(\mathbf{r},t) \cdot \hat{\mathbf{x}},\\
         \lim_{x \rightarrow 0^-} \mathbf{H}^{\mathrm{in}}(\mathbf{r},t) \times \hat{\mathbf{x}} &= \lim_{x \rightarrow 0^+} \mathbf{H}^{\mathrm{in}}(\mathbf{r},t) \times \hat{\mathbf{x}}.
     \end{align}
    To be consistent with our previous results, we work with the thickness-averaged fields, in which case the boundary conditions above can be recast in the form
    \begin{align}
         \lim_{x \rightarrow 0^-} B^{\mathrm{in}}_{\mathrm{ave},x}(\boldsymbol{\rho},t) &= \lim_{x \rightarrow 0^+} B^{\mathrm{in}}_{\mathrm{ave},x}(\boldsymbol{\rho},t),\\
         \lim_{x \rightarrow 0^-} H^{\mathrm{in}}_{\mathrm{ave},y}(\boldsymbol{\rho},t)  &= \lim_{x \rightarrow 0^+} H^{\mathrm{in}}_{\mathrm{ave},y}(\boldsymbol{\rho},t).
     \end{align}
    It should be noted that, although the $z$-axis is also parallel to the interface, we cannot use $ H^{\mathrm{in}}_{\mathrm{ave},z}(\boldsymbol{\rho},t)$ in the tangential boundary condition on the $\mathbf{H}$-field, for reasons that are explained in Appendix \ref{Appendix Reflection and Transmission Amplitudes}. Finally, neglecting the exchange field contribution to the magnetic (magnetizing) field, which is valid in the long-wavelength limit, and using the fact that the static field $\mathbf{B}_0$ ($\mathbf{H}_0$) is the same on both sides of the interface at $x=0$, we need only consider the contributions of the dipolar field, $\mathbf{B}^{\mathrm{in}}_{\mathrm{d,ave}}$ ($\mathbf{H}^{\mathrm{in}}_{\mathrm{d,ave}}$), and the superconductor-response field, $\mathbf{B}^{\mathrm{in}}_{\mathrm{sc, ave}}$ ($\mathbf{H}^{\mathrm{in}}_{\mathrm{sc,ave}}$) to the total field $\mathbf{B}^{\mathrm{in}}_{\mathrm{ave}}$ ($\mathbf{H}^{\mathrm{in}}_{\mathrm{ave}}$).

    The most general solution to the scattering problem is given in Appendix \ref{Appendix Reflection and Transmission Amplitudes}. Here, we present the result valid for the London dipolar limit, where $kd,k\lambda_L \ll 1$ and $d_s \gtrsim \lambda_L$. The coefficients $R_{12}$ and $T_{12}$ are given by
    \begin{align}
        R_{12}(\mathbf{k}) &= \dfrac{-\xi \bigg( \dfrac{2k_{T,x}}{k_T}-\dfrac{k_{x}}{k} \bigg)-1}{\xi \bigg( \dfrac{2k_{T,x}}{k_T}+\dfrac{k_{x}}{k} \bigg)+1}, \label{Equation First Fresnel Equation}\\
        T_{12}(\mathbf{k}) &= \dfrac{2\xi \dfrac{k_x}{k}}{\xi \bigg( \dfrac{2k_{T,x}}{k_T}+\dfrac{k_{x}}{k} \bigg)+1},
    \end{align}
    while for $R_{21}$ and $T_{21}$ we find
    \begin{align}
        R_{21}(\mathbf{k}) &= \dfrac{\xi \bigg( \dfrac{k_{T,x}}{k_T}-\dfrac{2k_{x}}{k} \bigg)-1}{\xi \bigg( \dfrac{2k_{R,x}}{k_R}-\dfrac{k_{T,x}}{k_T} \bigg)+1},\\
        T_{21}(\mathbf{k}) &= \dfrac{2\xi \bigg( \dfrac{k_{R,x}}{k_R}-\dfrac{k_{x}}{k} \bigg)}{\xi \bigg( \dfrac{2k_{R,x}}{k_R}-\dfrac{k_{T,x}}{k_T} \bigg)+1}. \label{Equation Last Fresnel Equation}
    \end{align}
    Equations~\eqref{Equation First Fresnel Equation}--\eqref{Equation Last Fresnel Equation} constitute the spin-wave analogue of the Fresnel equations encountered in classical optics \cite{BornWolf1999}. Plots of the reflection and transmission amplitudes as a function of the incident angle $\phi$ are given in Fig.~\ref{Figure Fresnel Amplitudes}. 

    \begin{figure*}[t]
    \centering
    \includegraphics[width=0.9\linewidth]{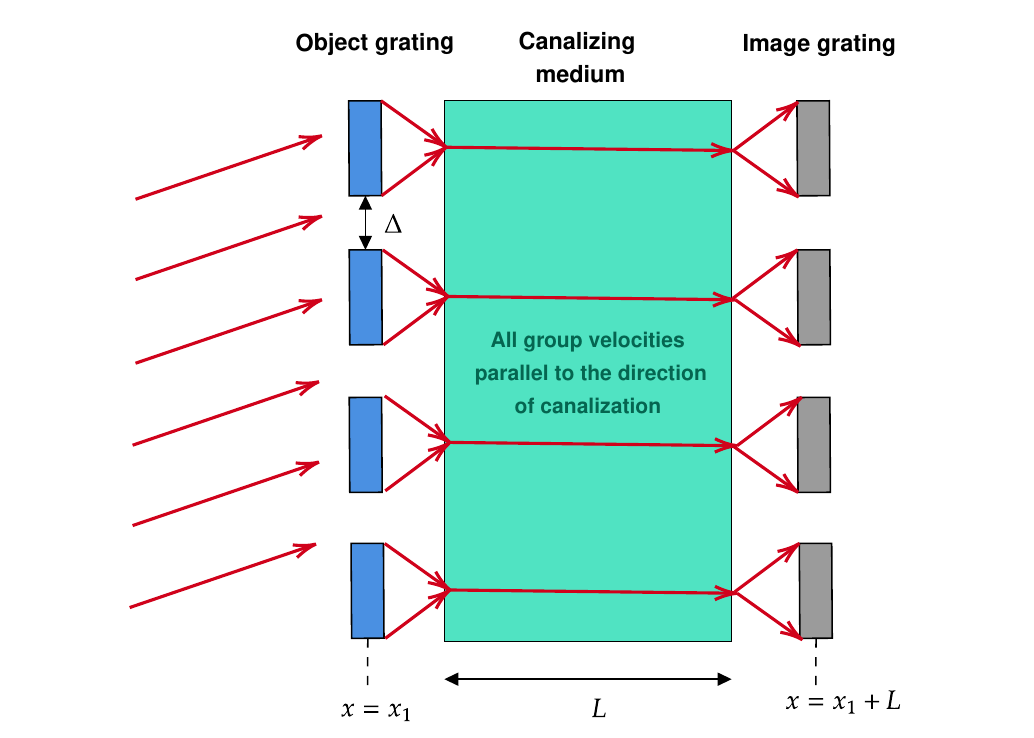}
    \caption{\textbf{Perfect imaging with a canalizing medium.} A grating (object) with spacing $\Delta$ is located at $x=x_1$ and illuminated by a spin wave, whose group velocity points in the direction of the red arrows. A canalizing medium of width $L$ is placed behind the grating, which causes the group velocity of the diffracted spin waves to align along a single direction. A perfect image of the grating is formed right behind the canalizing medium, at $x = x_1+L$, and this medium allows for features that are much smaller than the wavelength of the incident spin wave to be imaged without distortion. The FMI+SC heterostructure naturally behaves like a canalizing medium for spin waves.}
    \label{Figure Perfect Imaging}
\end{figure*}
    
    Before moving on to discuss several applications of FMI-SC heterostructures, we briefly comment on the properties of the reflection and transmission amplitudes. Firstly, we find that the amplitudes considered here behave similarly to the optical Fresnel amplitudes for $s$-polarized light \cite{Hecht2017,BornWolf1999}, because they satisfy relationships of the form
    \begin{align}
        T_{12}(\mathbf{k}) &= 1 + R_{12}(\mathbf{k}),\\
        T_{21}(\mathbf{k}) &= 1 + R_{21}(\mathbf{k}).
    \end{align}
    This can be seen directly from Fig.~\ref{Figure Fresnel Amplitudes}. Furthermore, we observe that the magnitude of the transmission amplitudes is typically much larger than that of the reflection amplitudes, which is consistent with experimental observations made in Ref.~\cite{Borst2023}. Finally, we observe a pronounced directional anisotropy in the reflection and transmission amplitudes between the cases where the incident spin wave has a positive $x$-component of the group velocity, Fig.~\ref{Figure Fresnel Amplitudes}(a), and where it has a negative $x$-component, Fig.~\ref{Figure Fresnel Amplitudes}(b), which can be understood immediately from the nonreciprocal nature of dipolar spin waves as discussed in Sect.~\ref{Section Reflection Refraction}.

\section{Applications}\label{Section Applications}
Here, we discuss several potential applications enabled by the unconventional reflection and refraction properties of the FMI-SC system, including perfect imaging and lensing of spin waves, waveguides and spin-wave interferometry. The aim is primarily to provide a qualitative overview, highlighting the potential advantages of this system over conventional approaches.

\subsection{Perfect imaging}
One of the main challenges in optics is overcoming the diffraction limit to achieve subwavelength imaging, i.e., resolving features smaller than the wavelength of the illuminating light. In conventional optical systems, however, the resolution is fundamentally limited by the Abbe diffraction limit \cite{BornWolf1999,Hecht2017}, preventing the recovery of fine spatial details. To overcome this limitation, Pendry showed that negative-index materials, characterized by simultaneously negative permittivity and permeability, can be used to construct superlenses capable of subwavelength imaging \cite{Pendry2000}. This capability arises from the unique response of these materials, where propagating waves are refocused through negative refraction, while evanescent waves are restored through their amplification inside the negative-index slab. The latter is associated with the excitation of surface plasmon modes at the slab interfaces \cite{Ramakrishna2002,GomezSantos2003}, which allows subwavelength information to be recovered.

Following Pendry’s proposal, several alternative approaches have been explored to achieve imaging beyond the diffraction limit. Among these, the concept of \textit{canalization} \cite{Belov2005,Belov2006,Belov2006_2} is particularly promising for spin-wave optics. Canalization is a wave-transport mechanism in which a medium is designed such that different spatial Fourier components of an image acquire the same phase during propagation, thereby suppressing diffraction and preserving subwavelength image information. This condition is realized when the dispersion relation exhibits sufficiently flat isofrequency contours, as we now show with a demonstrative calculation.

Let us consider a grating that is located at $x=x_1$ and illuminated by a plane wave with frequency $\omega$, as illusrated in Fig.~\ref{Figure Perfect Imaging}. We consider the image plane at $x=x_1+L$, with the region between the object (grating) and image planes occupied by a canalizing medium described by the dispersion relation $\omega = c k_x$, where $c\in\mathbb{R}_{>0}$ and $k_x>0$. The near-field distribution $\Psi(x,y)$ generated by the grating contains information about its spatial structure and can be decomposed into a spectrum of wavevector components,
\begin{align}
    \Psi(x,y)=\int dk_y\,\mathcal{F}[\Psi](k_y) e^{i(k_x(k_y)x+k_y y)},
\end{align}
where $\mathcal{F}[\Psi](k_y)$ is the spatial Fourier spectrum of the object and $k_x$ explicitly depends on $k_y$ via the dispersion relation. In a conventional medium, the dispersion is such that spatial components with large parallel wavevector components $k_y$ become evanescent during propagation, leading to the loss of subwavelength information. In contrast, a canalizing medium preserves these components by ensuring that the normal wavevector component $k_x$ remains independent of the parallel wavevector component $k_y$. Indeed, using $k_x = \omega/c$, we immediately have
\begin{align}\label{Equation Perfect Imaging}
    \Psi(x_1+L,y) &= e^{i(\omega/c)L} \int dk_y\,\mathcal{F}[\Psi](k_y)
    e^{i(k_xx_1+k_y y)} \nonumber\\
    &= e^{i(\omega/c)L}\Psi(x_1,y).
\end{align}
Thus, the spatial profile of the grating is reproduced at the image plane up to a global phase factor, which demonstrates that canalization can be used for \textit{perfect imaging}.

It should be noted that subwavelength imaging based on canalization (or self-collimation) has previously been demonstrated experimentally by exploiting the naturally occurring flat regions of the isofrequency contours of dipolar spin waves~\cite{Mansfeld2012,Makartsou2024}. However, the flattening is typically confined to a limited portion of the isofrequency contour, so that only a restricted range of wavevectors undergoes canalization. As a result, only part of the spatial Fourier spectrum of the image can be faithfully transmitted. In contrast, our results show that Meissner screening by a superconducting overlayer produces isofrequency contours that are substantially flatter/straighter and remain flat over a much broader range of wavevectors than in conventional dipolar spin-wave systems. Consequently, a much larger portion of the spatial spectrum can be transmitted without diffraction, making the FMI-SC system significantly better suited for subwavelength imaging based on canalization.

\subsection{Lenses}
In recent years, there has been increasing interest in the development of spin-wave lenses \cite{Toedt2016,Whitehead2018,Grafe2020,Bao2020,Dai2022}. Due to the unconventional refraction properties of the FMI-SC system, including regimes with negative phase and group velocity refraction, we expect it to be a promising platform for the realization of a variety of spin-wave lensing functionalities.

Here, we briefly consider a particular type of lens that can be constructed by exploiting the strong temperature dependence of the refraction properties of the FMI-SC system. In optics, gradient-index (GRIN) lenses are of particular interest because they enable focusing through a spatial variation of the refractive index rather than through curved interfaces  \cite{Hecht2017}. This allows for the realization of lenses with flat surfaces and can reduce optical aberrations compared to conventional lenses \cite{Hecht2017}. As shown in Section~\ref{Section Reflection Refraction}, the reflection and refraction properties of the FMI-SC system can be tuned strongly by varying the temperature. Introducing a temperature gradient therefore provides a direct route towards creating a GRIN spin-wave lens, where the local refraction properties vary continuously across the system.

\subsection{Waveguides}
While most spin-wave waveguides rely on geometrical confinement or engineered magnetic textures \cite{Demidov2009,Wagner2016,Henry2019,Gallardo2022}, canalization (or self-collimation) provides a fundamentally different guiding mechanism based on anisotropic dispersion \cite{Schneider2010}. As a consequence of the straight isofrequency contours in the FMI-SC system, the group velocity direction becomes nearly independent of the wavevector within the canalization regime. Therefore, spin-wave beams corresponding to different $\mathbf{k}$-vectors propagate in the same direction, enabling robust beam transport and steering. The direction of propagation can be directly controlled by varying the saturation magnetization angle $\psi$, thus providing a simple mechanism to tune the waveguide orientation.

An additional advantage of the FMI-SC system is that Meissner screening by the superconductor leads not only to flatter isofrequency contours but also to a significant enhancement of the group velocity compared to conventional dipolar spin-wave systems \cite{Zhou2024}. This enables faster propagation of spin-wave signals, which is advantageous for applications requiring rapid information transfer over spin-wave channels.

\subsection{Interferometry}
Spin-wave interferometry has emerged as an important building block for coherent magnonic devices \cite{Rousseau2015,Kanazawa2016,Papp2021}.
The unusual reflection and refraction properties of spin waves in the FMI–SC system naturally lend themselves to several interferometric functionalities, which we now discuss in some detail.

One useful property of the FMI–SC system is the strong temperature dependence of its reflection and refraction characteristics, including the ability to open or completely suppress specific reflection channels. This suggests the possibility of temperature-controlled spin-wave mirrors and beam splitters, where the amplitudes and directions of reflected and transmitted spin waves can be modified through temperature. Such elements could serve as building blocks for Mach–Zehnder- or Michelson-type magnonic interferometers, where the splitting ratio and resulting interference pattern can be tuned without changing the geometry of the device.

Another essential element for interferometric applications is a controllable phase shifter, which can also be realized using the FMI–SC system. Unlike conventional phase shifters, where the accumulated phase generally increases with increasing path length, negative phase refraction in the FMI–SC system can also enable a reduction of the accumulated phase. Furthermore, since Meissner screening strongly modifies only one branch of the spin-wave dispersion, the resulting phase shift is highly dependent on the propagation direction. Spin waves incident on the phase-shifting region from opposite directions can therefore experience completely different phase modifications. Finally, as shown in Eq.~\eqref{Equation Perfect Imaging}, for an appropriate choice of $\psi$, a phase shifter can be constructed for which spin waves propagating along the modified dispersion branch acquire the same phase shift regardless of their incident direction. This property may be useful for designing spin-wave interferometric elements where a well-defined phase relation between different propagation paths is required.

More broadly, the above discussion indicates that the FMI–SC system can be used to realize several key elements required for spin-wave interferometry, including tunable mirrors, beam splitters, and phase-shifting elements, with the strong temperature dependence of the system providing an additional control parameter for tuning the interference conditions without modifying the device geometry.

\section{Conclusion}
In this work, we have developed a general framework for spin-wave optics in ferromagnetic insulator–superconductor (FMI–SC) heterostructures and shown that superconducting Meissner screening gives rise to a range of unconventional spin-wave phenomena. In particular, we demonstrated negative reflection, negative phase- and group-velocity refraction, and reflection and refraction laws that differ fundamentally from their optical counterparts. We further derived the spin-wave Fresnel equations governing reflection and transmission at FMI–SC interfaces and showed that the scattering properties can be controlled through temperature.

Perhaps most importantly, we have demonstrated that Meissner screening can flatten one branch of the dipolar spin-wave dispersion far beyond what is achievable in conventional dipolar systems. These nearly straight isofrequency contours enable functionalities including perfect spin-wave imaging, efficient waveguiding, and interferometric elements, such as phase shifters and beam splitters, with unconventional operating properties.

It should be emphasized that, for applications such as spin-wave imaging/lensing and interferometry, engineering the desired dispersion relation is only one part of the design process. The reflection and transmission coefficients, through both their amplitudes and phases, also play a central role in determining the resulting spin-wave field. In the present work, we have focused on the reflection and transmission coefficients at the interface between a bare FMI and an FMI-SC heterostructure. For specific applications, however, it may be desirable to engineer the interface itself to tailor these coefficients, for example by modifying its geometry or introducing intermediate matching layers.

In this paper, we have exclusively treated superconductivity on the level of London theory. For sufficiently strong magnetic fields, however, phenomena beyond the London description can emerge, including the formation of Abrikosov-vortices in type-II superconductors \cite{Abrikosov1957,Tinkham1996}. Experimental and theoretical studies have revealed that ordered Abrikosov vortex lattices can interact strongly with magnons \cite{Bespalov2014,Dobrovolskiy2022,Niedzielski2023,Katkov2024,Dobrovolskiy2025}, leading to significant modifications of spin-wave dispersion and propagation properties. Such effects could therefore be of substantial interest for spin-wave optics applications, potentially enabling functionalities beyond those proposed in this work.

Another interesting direction for future work is to investigate the influence of additional magnetic interactions on the spin-wave dispersion. Beyond the dipolar and exchange interactions considered here, one could include the effects of uniaxial or crystalline anisotropies, as well as the Dzyaloshinskii--Moriya interaction. Although these interactions were not the focus of the present work, numerical calculations indicate that the nearly straight isofrequency contours, as well as phenomena such as negative reflection and refraction, remain robust provided that the dipolar and Zeeman contributions continue to dominate the spin-wave dispersion.

Finally, the results presented here establish FMI–SC heterostructures as a versatile platform for tunable spin-wave optics and provide new opportunities for engineering magnonic devices based on superconducting proximity effects. Among the potential applications discussed here, we considered perfect imaging devices, spin-wave lenses, waveguides and interferometric elements, such as spin-wave beam splitters and phase shifters. Beyond the experimental realization of these devices, we expect that many intriguing results can be obtained by studying spin-wave analogues of optical phenomena in this system, potentially enabling novel spintronic applications.

\section{Acknowledgements}
This project has received funding from the European Research Council (ERC) under the European Union’s Horizon 2020 research and innovation programme (grant agreement No 101170480 MAGICWAVE) and from by the Dutch Research Council (NWO) under awards VI.Vidi.193.077, NGF.1582.22.018, and OCENW.XL21.XL21.058.

	\newpage
	\appendix
	\onecolumngrid
	
	\section{Integral identities}\label{Appendix Integral Identities}
	Here, we derive several integral identities of which we make frequent use in our calculations. These include the dipolar field integrals and the integrals needed to verify the ansatz for the magnetic-field solution inside the superconductor.

	\subsection{Dipolar field integrals}

    \subsubsection{Infinite magnetic film}
    We first consider a magnetic film that extends to $\pm\infty$ in both the $x$- and $y$-directions. Starting from a magnetization density $	\mathbf{m}_{\mathbf{k}}(\mathbf{r},t)$ given by
	\begin{align}
		\mathbf{m}_{\mathbf{k}}(\mathbf{r},t) &= e^{i (\mathbf{k}\cdot \boldsymbol{\rho}-\omega t)} \bigg[ \theta(z+d/2)-\theta(z-d/2) \bigg]
		\begin{pmatrix}
			i m_x\\i m_y\\m_z
		\end{pmatrix}
		,
	\end{align}
	we seek to derive the resulting dipolar (magnetizing) field $\mathbf{H}_d(\mathbf{r},t)$. Since no free current density is present and the time variation of the electric displacement field is expected
	to be small, this field can be calculated from the magnetic scalar potential $\Phi_{\mathbf{k}}(\mathbf{r},t)$, yielding \cite{Jackson1999}
	\begin{align}
		H_{\mathrm{d}}(\mathbf{r},t) &= \nabla \Phi_{\mathbf{k}}(\mathbf{r},t) ,
	\end{align}
    with
    \begin{align}
        \Phi_{\mathbf{k}}(\mathbf{r},t) &=\frac{1}{4\pi} \int d^3 \mathbf{r}'\, \frac{\nabla' \cdot \mathbf{m}_{\mathbf{k}}(\mathbf{r}',t)}{|\mathbf{r}-\mathbf{r}'|}
    \end{align}
	For the problem at hand, $\Phi_{\mathbf{k}}(\mathbf{r},t)$ is given by
	\begin{align}
		\Phi_{\mathbf{k}} (\mathbf{r},t) &= \frac{1}{4\pi} \int d^3 \mathbf{r}' \frac{e^{i (\mathbf{k}\cdot \boldsymbol{\rho}'-\omega t)}\mathcal{G}_{\mathbf{k}}(z')}{|\mathbf{r}-\mathbf{r}'|},
	\end{align}
	where
	\begin{align}
		\mathcal{G}_{\mathbf{k}}(z) = - \mathbf{k} \cdot \mathbf{m} \bigg[ \theta(z+d/2)-\theta(z-d/2) \bigg] + m_z \bigg[ \delta(z+d/2)-\delta(z-d/2) \bigg],
	\end{align}
    and with $\mathbf{m} = (m_x,m_y,m_z)$. Now, making use of a Fourier identity \cite{Jackson1999},
	\begin{align}\label{Equation Fourier Transform Coulomb Term}
		\frac{1}{|\mathbf{\mathbf{r}-\mathbf{r}'}|} &= \frac{1}{2 \pi}\int d^2 \mathbf{q}\, \frac{e^{-q|z-z'|}}{q} e^{i \mathbf{q}\cdot (\boldsymbol{\rho}-\boldsymbol{\rho}')}
	\end{align}
	we immediately have
	\begin{align}
		\Phi_{\mathbf{k}}(\mathbf{r},t) &= \frac{1}{8\pi^2} \int d^2 \mathbf{q} \int d^3 \mathbf{r}' \frac{e^{i (\mathbf{k}\cdot \boldsymbol{\rho}'-\omega t)}\mathcal{G}_{\mathbf{k}}(z')}{q} e^{-q|z-z'|} e^{i \mathbf{q}\cdot (\boldsymbol{\rho}-\boldsymbol{\rho}')} \nonumber\\
		&= \frac{1}{2 k} e^{i (\mathbf{k}\cdot \boldsymbol{\rho}-\omega t)}  \int d z' e^{-k|z-z'|}\mathcal{G}_{\mathbf{k}}(z').\label{Equation IR3}
	\end{align}
	Here, we have used the Dirac delta identity given by
	\begin{align}
		\frac{1}{2\pi} \int_{-\infty}^{\infty} d x \, e^{i k x} = \delta(k).
	\end{align}
	Now, to complete the calculation we need to evaluate the following integral over $z'$,
	\begin{align}
		\bar{\mathcal{G}}_{\mathbf{k}}(z) = \int d z' e^{-k|z-z'|}\mathcal{G}_{\mathbf{k}}(z'),
	\end{align}
	where we distinguish between the cases $|z| < d/2$ and $|z|>d/2$. We find
	\begin{align}
		\bar{\mathcal{G}}_{\mathbf{k}}(z) =
		\begin{cases}
			 \dfrac{2\, \mathbf{k} \cdot \mathbf{m}}{k}  \left[ e^{-kd/2}\cosh\left(k z \right)-1 \right]-2 m_z e^{-kd/2} \sinh \left( kz \right), \hspace{0.1cm} &\text{for} \hspace{0.1cm} |z|<d/2,\\
			- 2 \sinh \left( \dfrac{k d}{2} \right) \left[ \dfrac{\mathbf{k} \cdot \mathbf{m}}{k} + \text{sgn}(z) m_z \right] e^{-k|z|},  \hspace{0.1cm} &\text{for} \hspace{0.1cm} |z|>d/2.
		\end{cases}
	\end{align}
	In a straightforward manner, we then find that the dipolar magnetizing field is given by
	\begin{align}
		\mathbf{H}_d(\mathbf{r},t) &= \frac{1}{2 k} e^{i (\mathbf{k} \cdot \boldsymbol{\rho}-\omega t)}
		\begin{pmatrix}
			i k_x \bar{\mathcal{G}}_{\mathbf{k}}(z)\vspace{0.1cm}\\ 
			i k_y \bar{\mathcal{G}}_{\mathbf{k}}(z) \vspace{0.1cm}\\
			\bar{\mathcal{G}}_{\mathbf{k}}'(z)\vspace{0.1cm}
		\end{pmatrix}
		,
	\end{align}
	where
	\begin{align}
		\bar{\mathcal{G}}_{\mathbf{k}}'(z) =
		\begin{cases}
			2 \, \mathbf{k} \cdot \mathbf{m} e^{-kd/2} \sinh\left(k z \right)- 2 k \,m_z e^{-kd/2}\cosh \left( kz \right), \hspace{0.1cm} &\text{for} \hspace{0.1cm} |z|<d/2,\\
			- \text{sgn}(z) k \,\bar{\mathcal{G}}_{\mathbf{k}}(z),  \hspace{0.1cm} &\text{for} \hspace{0.1cm} |z|>d/2.
		\end{cases}
	\end{align}
    Finally, in our calculation of the dispersion relation, we work with the \textit{average} dipolar field \textit{inside} the magnetic film, which is given by
    \begin{align}
        \mathbf{H}^{\mathrm{in}}_{\mathrm{d,ave}}(\boldsymbol{\rho},t) &= \frac{1}{d} \int_{-d/2}^{d/2} dz \, \mathbf{H}_d(\mathbf{r},t) \nonumber \\
        &= \dfrac{e^{i (\mathbf{k} \cdot \boldsymbol{\rho}-\omega t)}}{k^2}
        \begin{pmatrix}
            i k_x (\mathbf{k} \cdot \mathbf{m})[g(k)-1]\\
            i k_y (\mathbf{k} \cdot \mathbf{m})[g(k)-1]\\
            -k^2\,m_z\,g(k)
        \end{pmatrix}
        ,
    \end{align}
    where $g(k)$ is defined as
    \begin{align}
        g(k) &= \dfrac{1}{k d} \left(1-e^{-k d} \right).
    \end{align}

	\subsection{Ansatz field integrals}
	We assume the following ansatz for the magnetic-field solution inside the superconducting region $\Omega_S$,
	\begin{align}
		\mathbf{B}(\mathbf{r},t) &= e^{i (\mathbf{k}\cdot \boldsymbol{\rho}-\omega t)}
		\begin{pmatrix}
			i k_x \bigg[ C_1 e^{-\zeta z}+C_2 e^{\zeta z} \bigg] \vspace{0.2cm}\\
			i k_y \bigg[ C_1 e^{-\zeta z}+C_2 e^{\zeta z} \bigg]\vspace{0.2cm}\\
			-\frac{k^2}{\zeta}C_1 e^{-\zeta z}+\frac{k^2}{\zeta}C_2 e^{\zeta z}+C_3
		\end{pmatrix}
		,
	\end{align}
	where $z \in [d/2,d/2+d_s]$. Referring to Eq.~\eqref{Equation Self-Consistent Equation}, we are interested in integrals of the form
	\begin{align}
		\mathbf{I}_{\Omega_S}(\mathbf{r},t) &= \int_{\Omega_S} d^3 \mathbf{r}'\biggr[\frac{\mathbf{B}(\mathbf{r}',t)}{|\mathbf{r}-\mathbf{r}'|}\biggr], \\ \mathbf{I}_{\partial\Omega_S}(\mathbf{r},t) &= \oint_{\partial \Omega_S} d^2 \mathbf{r}' \frac{(\nabla' \times \mathbf{B}(\mathbf{r}',t))\times \hat{\mathbf{n}}}{|\mathbf{r}-\mathbf{r}'|}.
	\end{align}
	Focusing first on $\mathbf{I}_{\Omega_S}$, we directly find from the Fourier identity of Eq.~\eqref{Equation Fourier Transform Coulomb Term} that
	\begin{align}
		\int_{\Omega_S} d^3 \mathbf{r}' \frac{e^{i (\mathbf{k}\cdot \boldsymbol{\rho}' -\omega t)}e^{\pm \zeta z'}}{|\mathbf{r}-\mathbf{r}'|} &= \frac{2 \pi}{k}e^{i (\mathbf{k}\cdot \boldsymbol{\rho} -\omega t)} \biggr[ \frac{2 k}{k^2-\zeta^2} e^{\pm \zeta z}-e^{\pm \zeta d/2} \biggr( \frac{e^{k d/2}}{k \pm \zeta}e^{-kz}+\frac{e^{-k d/2}e^{(-k\pm \zeta)d_s}}{k \mp \zeta}e^{kz}\biggr)\biggr],
	\end{align}
	for $z \in [d/2,d/2+d_s]$, while for $z \in [-d/2,d/2]$ we have
	\begin{align}
		\int_{\Omega_S} d^3 \mathbf{r}' \frac{e^{i (\mathbf{k}\cdot \boldsymbol{\rho}' -\omega t)}e^{\pm \zeta z'}}{|\mathbf{r}-\mathbf{r}'|} &= \frac{2 \pi}{k}e^{i (\mathbf{k}\cdot \boldsymbol{\rho} -\omega t)} e^{k z} \biggr[ \frac{e^{(-k\pm\zeta)d/2}}{k \mp \zeta}  \biggr(1-e^{(-k\pm\zeta)d_s} \biggr)   \biggr].
	\end{align}
	From these results, we then immediately have
	\begin{align}\label{Equation Solution I Omega S inside SC}
		\mathbf{I}_{\Omega_S}(\mathbf{r},t) e^{-i (\mathbf{k}\cdot \boldsymbol{\rho}-\omega t)}  &= \frac{4 \pi}{k^2} C_3
		\begin{pmatrix}
			0\\0\\1
		\end{pmatrix}
		+
		\frac{4 \pi}{k^2-\zeta^2} C_1 \,e^{-\zeta z}
		\begin{pmatrix}
			i k_x\\
			i k_y\\
			-k^2/\zeta
		\end{pmatrix}
		+\frac{4 \pi}{k^2-\zeta^2} C_2 \,e^{\zeta z}
		\begin{pmatrix}
			i k_x\\
			i k_y\\
			k^2/\zeta
		\end{pmatrix}
		\nonumber \\
		&-\frac{2 \pi\, e^{k d/2}}{k}e^{- k z}
		\begin{pmatrix}
			i k_x \bigg[ C_1 \frac{e^{-\zeta d/2}}{k-\zeta}+C_2 \frac{e^{\zeta d/2}}{k+\zeta} \bigg]\vspace{0.2cm}\\
			i k_y \bigg[ C_1 \frac{e^{-\zeta d/2}}{k-\zeta}+C_2 \frac{e^{\zeta d/2}}{k+\zeta} \bigg]\vspace{0.2cm}\\
			\frac{k^2}{\zeta} \bigg[ -C_1 \frac{e^{-\zeta d/2}}{k-\zeta}+C_2 \frac{e^{\zeta d/2}}{k+\zeta}+ \frac{\zeta}{k^3}C_3 \bigg]
		\end{pmatrix}
		\nonumber \\
		&-\frac{2 \pi\, e^{-k (d/2+d_s)}}{k}e^{k z}
		\begin{pmatrix}
			i k_x \bigg[ C_1 \frac{e^{-\zeta (d/2+d_s)}}{k+\zeta}+C_2 \frac{e^{\zeta (d/2+d_s)}}{k-\zeta} \bigg]\vspace{0.2cm}\\
			i k_y \bigg[ C_1 \frac{e^{-\zeta (d/2+d_s)}}{k+\zeta}+C_2 \frac{e^{\zeta (d/2+d_s)}}{k-\zeta} \bigg]\vspace{0.2cm}\\
			\frac{k^2}{\zeta} \bigg[ -C_1 \frac{e^{-\zeta (d/2+d_s)}}{k+\zeta}+C_2 \frac{e^{\zeta (d/2+d_s)}}{k-\zeta}- \frac{\zeta}{k^3}C_3 \bigg]
		\end{pmatrix}
		,
	\end{align}
	for $z \in [d/2,d/2+d_s]$, while for $z \in [-d/2,d/2]$ we have
	\begin{align}
		\mathbf{I}_{\Omega_S}(\mathbf{r},t) e^{-i (\mathbf{k}\cdot \boldsymbol{\rho}-\omega t)}  &=
		\dfrac{2 \pi e^{-k (d/2+d_s)}}{k} e^{k z}
		\begin{pmatrix}
			&i k_x \bigg[ C_1 \frac{e^{-\zeta d/2}}{k + \zeta}  \biggr(e^{k d_s}-e^{-\zeta d_s} \biggr) + C_2 \frac{e^{\zeta d/2}}{k - \zeta}  \biggr(e^{k d_s}-e^{\zeta d_s} \biggr) \bigg] \vspace{0.2cm}\\
			&i k_y \bigg[ C_1 \frac{e^{-\zeta d/2}}{k + \zeta}  \biggr(e^{k d_s}-e^{-\zeta d_s} \biggr) + C_2 \frac{e^{\zeta d/2}}{k - \zeta}  \biggr(e^{k d_s}-e^{\zeta d_s} \biggr) \bigg]    \vspace{0.2cm}\\
			&k^2/\zeta\bigg[ -C_1 \frac{e^{-\zeta d/2}}{k + \zeta}  \biggr(e^{k d_s}-e^{-\zeta d_s} \biggr) + C_2 \frac{e^{\zeta d/2}}{k - \zeta}  \biggr(e^{k d_s}-e^{\zeta d_s} \biggr) \bigg] + C_3 \dfrac{e^{k d_s}-1}{k}
		\end{pmatrix}
		.
	\end{align}
	
	Having determined $\mathbf{I}_{\Omega_S}$, we now switch our attention to $\mathbf{I}_{\partial\Omega_S}$. Since the other boundary planes of $\partial \Omega_S$ are infinitely far away, we need only to consider the planes $z = d/2$ and $z = d/2 + d_s$. Now, because 
	\begin{align}
		\bigg( \nabla \times \mathbf{B}(\mathbf{r},t) \bigg) \times \hat{\mathbf{z}} &= i e^{i (\mathbf{k}\cdot \boldsymbol{\rho}-\omega t)} \bigg[ \bigg(-\zeta+\frac{k^2}{\zeta} \bigg)C_1 e^{- \zeta z} + \bigg(\zeta-\frac{k^2}{\zeta} \bigg)C_2 e^{ \zeta z}-C_3 \bigg] \mathbf{k},
	\end{align}
	we immediately find
	\begin{align}\label{Equation Solution I partial Omega S}
		\mathbf{I}_{\partial\Omega_S}(\mathbf{r},t) e^{-i (\mathbf{k}\cdot \boldsymbol{\rho}-\omega t)} &=  \frac{2\pi}{k}\bigg( \zeta - \frac{k^2}{\zeta} \bigg) e^{-k|z-d/2|}
		\begin{pmatrix}
			i k_x \bigg(C_1 e^{-\zeta d/2}-C_2 e^{\zeta d/2} - \frac{\zeta}{k^2-\zeta^2} C_3 \bigg)\vspace{0.2cm}\\ 
			i k_y \bigg(C_1 e^{-\zeta d/2}-C_2 e^{\zeta d/2} - \frac{\zeta}{k^2-\zeta^2} C_3 \bigg) \vspace{0.2cm}\\ 
			0
		\end{pmatrix}
		\nonumber\\
		&-\frac{2\pi}{k}\bigg( \zeta - \frac{k^2}{\zeta} \bigg) e^{-k|z-d/2-d_s|}
		\begin{pmatrix}
			i k_x \bigg(C_1 e^{-\zeta (d/2+d_s)}-C_2 e^{\zeta (d/2+d_s)} - \frac{\zeta}{k^2-\zeta^2} C_3 \bigg)\vspace{0.2cm}\\ 
			i k_y \bigg(C_1 e^{-\zeta (d/2+d_s)}-C_2 e^{\zeta (d/2+d_s)} - \frac{\zeta}{k^2-\zeta^2} C_3 \bigg) \vspace{0.2cm}\\ 
			0
		\end{pmatrix}
		,
	\end{align}
	for all $z \in(-\infty,\infty)$.

	\section{Derivation of $C$- and $\zeta$-coefficients}\label{Appendix Derivation of C, and kappa Coefficients}
	Here, we determine the coefficients $C_1,C_2,C_3 \in \mathbb{C}$ and $\zeta \in \mathbb{R}_{>0}$ of the magnetic-field ansatz given by
	\begin{align}
		\mathbf{B}(\mathbf{r},t) &= e^{i (\mathbf{k}\cdot \boldsymbol{\rho}-\omega t)}
		\begin{pmatrix}
			i k_x \bigg[ C_1 e^{-\zeta z}+C_2 e^{\zeta z} \bigg] \vspace{0.2cm}\\
			i k_y \bigg[ C_1 e^{-\zeta z}+C_2 e^{\zeta z} \bigg]\vspace{0.2cm}\\
			-\frac{k^2}{\zeta}C_1 e^{-\zeta z}+\frac{k^2}{\zeta}C_2 e^{\zeta z}+C_3
		\end{pmatrix}
		,
	\end{align}
	where $z \in [d/2,d/2+d_s]$. These coefficients are determined by plugging this solution into the self-consistent equation, Eq.~\eqref{Equation Self-Consistent Equation}, which, upon using the notation introduced in Appendix \ref{Appendix Integral Identities}, can be written as
	\begin{align}
		\mathbf{B}(\mathbf{r},t) &= \mathbf{B}_d(\mathbf{r},t)-\frac{1}{4\pi \lambda_L^2} \mathbf{I}_{\Omega_S}(\mathbf{r},t)+\frac{1}{4\pi} \mathbf{I}_{\partial\Omega_S}(\mathbf{r},t).
	\end{align}
	The solutions for $\mathbf{I}_{\Omega_S}(\mathbf{r},t)$ and $\mathbf{I}_{\partial \Omega_S}(\mathbf{r},t)$ inside the superconducting region are given by Eq.~\eqref{Equation Solution I Omega S inside SC} and \eqref{Equation Solution I partial Omega S}, respectively. Now, focusing on the $z$-component of the above equation, we arrive at the following linear system in $C_1$, $C_2$, $C_3$ and $\zeta$:
	\begin{align}
		\left\{
		\begin{aligned}
			&\zeta = \sqrt{k^2+\frac{1}{\lambda_L^2}},\\
			&B_{\mathrm{d}}(\mathbf{k}) + \frac{k\, e^{k d/2}}{2 \zeta \lambda_L^2} \biggr( C_1 \frac{e^{-\zeta d/2}}{k-\zeta}-C_2 \frac{e^{\zeta d/2}}{k+\zeta} \biggr) = 0,\\
			&C_1 \frac{e^{-\zeta (d/2+d_s)}}{k+\zeta}-C_2 \frac{e^{\zeta (d/2+d_s)}}{k-\zeta} =0,\\
			&C_3  = 0.
		\end{aligned}
		\right.
	\end{align}
	Here, we have introduced the term $B_{\mathrm{d}}(\mathbf{k})$, which is given by
	\begin{align}
		B_{\mathrm{d}}(\mathbf{k}) &= - \mu_0\sinh \left( \frac{k d}{2} \right) \left[ \frac{k_x}{k} m_x +  m_z \right]. 
	\end{align}
	We can now straightforwardly solve for $C_1$ and $C_2$, and we then arrive at the following set of solutions,
	\begin{align}
		\left\{
		\begin{aligned}
			\zeta &= \sqrt{k^2+\frac{1}{\lambda_L^2}},\\
			C_1 &= \frac{2 \zeta \, B_{\mathrm{d}}(\mathbf{k}) e^{(\zeta-k)d/2}}{a_{k,+}k^2}\frac{1}{1-\big(\frac{a_{k,-}}{a_{k,+}}\big)^2 e^{-2 \zeta d_s}},\\
			C_2&= \frac{2 a_{k,-}\zeta \, B_{\mathrm{d}}(\mathbf{k})e^{-(\zeta+k)d/2}}{a_{k,+}^2k^2}\frac{e^{-2 \zeta d_s}}{1-\big(\frac{a_{k,-}}{a_{k,+}}\big)^2 e^{-2 \zeta d_s}},\\
			C_3  &= 0,
		\end{aligned}
		\right.
	\end{align}
	where $a_{k,\pm} = 1 \pm \zeta/k$. It is straightforward to verify that we would have arrived at the same set of solutions had we started from the $x$- and $y$-components of the self-consistent equation. 

\section{Derivation of reflection and transmission coefficients}\label{Appendix Reflection and Transmission Amplitudes}
Here, we derive the reflection and transmission coefficients for the scattering problem introduced in Section \ref{Section Fresnel Equations}. These coefficients ($R_{12}$, $R_{21}$, $T_{12}$ and $T_{21}$) are defined via the magnetization densities $\mathbf{m}_{12}(\mathbf{k},\boldsymbol{\rho}) e^{-i\omega t}$ and $\mathbf{m}_{21}(\mathbf{k},\boldsymbol{\rho})e^{-i\omega t}$, where we recall that
\begin{align}
    \mathbf{m}_{12}(\mathbf{k},\boldsymbol{\rho}) &=
        \begin{cases}
        e^{i\mathbf{k}\cdot\boldsymbol{\rho}}\,
        \begin{pmatrix}
            i \eta_0(\mathbf{k}) \\ 0 \\ 1
        \end{pmatrix} 
        + R_{12}
        e^{i\mathbf{k}_{ R}\cdot\boldsymbol{\rho}}\,
        \begin{pmatrix}
            i \eta_0(\mathbf{k}_R) \\ 0 \\ 1
        \end{pmatrix} \hspace{2.2cm} (x < 0),\\ T_{12}
        e^{i\mathbf{k}_{T}\cdot\boldsymbol{\rho}}\,
        \begin{pmatrix}
            i \eta(\mathbf{k}_T) \\ 0 \\ 1
        \end{pmatrix}\hspace{5.15cm} (x > 0),
        \end{cases}
\end{align}
and
\begin{align}
        \mathbf{m}_{21}(\mathbf{k},\boldsymbol{\rho}) &=
        \begin{cases}
        T_{21}
        e^{i\mathbf{k}_{T}'\cdot\boldsymbol{\rho}}\,
        \begin{pmatrix}
            i \eta_0(\mathbf{k}_T') \\ 0 \\ 1
        \end{pmatrix} \hspace{5.2cm} (x < 0),\\ e^{i\mathbf{k}'\cdot\boldsymbol{\rho}}\,
        \begin{pmatrix}
            i \eta(\mathbf{k}') \\ 0 \\ 1
        \end{pmatrix} 
        + R_{21}
        e^{i\mathbf{k}_{ R}'\cdot\boldsymbol{\rho}}\,
        \begin{pmatrix}
            i \eta(\mathbf{k}_R') \\ 0 \\ 1
        \end{pmatrix}
        \hspace{2.5cm} (x > 0),
        \end{cases}
    \end{align}
for $-d/2 \leq z \leq d/2$. To determine the coefficients, we note that the solution to the scattering problem is required to satisfy magnetostatic boundary conditions. Denoting the total magnetic field and the total magnetizing field inside the film by $\mathbf{B}^{\mathrm{in}}$ and $\mathbf{H}^{\mathrm{in}}$, respectively, these boundary conditions are given by
     \begin{align}
         \lim_{x \rightarrow 0^-} \mathbf{B}^{\mathrm{in}}(\mathbf{r},t) \cdot \hat{\mathbf{x}} &= \lim_{x \rightarrow 0^+} \mathbf{B}^{\mathrm{in}}(\mathbf{r},t) \cdot \hat{\mathbf{x}},\\
         \lim_{x \rightarrow 0^-} \mathbf{H}^{\mathrm{in}}(\mathbf{r},t) \times \hat{\mathbf{x}} &= \lim_{x \rightarrow 0^+} \mathbf{H}^{\mathrm{in}}(\mathbf{r},t) \times \hat{\mathbf{x}}.
     \end{align}
Neglecting the exchange field contribution to the magnetic (magnetizing) field, which is valid in the long-wavelength limit, and using the fact that the static field $\mathbf{B}_0$ ($\mathbf{H}_0$) is the same on both sides of the interface at $x=0$, we need only consider the contributions of the dipolar field, $\mathbf{B}^{\mathrm{in}}_{\mathrm{d}}$ ($\mathbf{H}^{\mathrm{in}}_{\mathrm{d}}$), and the superconductor-response field, $\mathbf{B}^{\mathrm{in}}_{\mathrm{sc}}$ ($\mathbf{H}^{\mathrm{in}}_{\mathrm{sc}}$) to the total field $\mathbf{B}^{\mathrm{in}}$ ($\mathbf{H}^{\mathrm{in}}$).

To be consistent with our previous results, we would like to work with the thickness-averaged fields. However, one needs to be a bit careful here, for only the $x$- and $y$-components of these fields can be used, as we will now demonstrate. First, we note that the absence of a free current allows us to derive the dipolar field and the superconductor-response field inside the film using a magnetic scalar potential $\Phi(\mathbf{r},t)$, i.e., we can write $\mathbf{H}^{\mathrm{in}}(\mathbf{r},t) = \nabla \Phi(\mathbf{r},t)$. It then immediately follows that the thickness-averaged $\mathbf{H}$-field is given by
\begin{align}
    \mathbf{H}^{\mathrm{in}}_{\mathrm{ave}}(\boldsymbol{\rho},t) &= \dfrac{1}{d} \int_{-d/2}^{d/2} dz \, \mathbf{H}^{\mathrm{in}}(\mathbf{r},t) \nonumber\\&= \dfrac{1}{d} \int_{-d/2}^{d/2} dz \,\nabla \Phi(\mathbf{r},t).
\end{align}
Now, introducing the thickness-averaged magnetic scalar potential,
\begin{align}
    \Phi_{\mathrm{ave}}(\boldsymbol{\rho},t) & = \dfrac{1}{d} \int_{-d/2}^{d/2} dz \, \Phi(\mathbf{r},t),
\end{align}
we see that the previous equation can also be written as
\begin{align}\label{Equation Invalidity of z-component Appendix}
    \mathbf{H}^{\mathrm{in}}_{\mathrm{ave}}(\boldsymbol{\rho},t) &= 
    \begin{pmatrix}
        \partial_x \Phi_{\mathrm{ave}}(\boldsymbol{\rho},t)\\
        \partial_y \Phi_{\mathrm{ave}}(\boldsymbol{\rho},t)\\
        \dfrac{1}{d} \bigg[ \Phi(\mathbf{r},t)\bigg|_{z = d/2} - \Phi(\mathbf{r},t)\bigg|_{z = -d/2}  \bigg]
    \end{pmatrix}
    .
\end{align}
From Eq.~\eqref{Equation Invalidity of z-component Appendix}, we conclude that the $x$- and $y$-components of $\mathbf{H}^{\mathrm{in}}_{\mathrm{ave}}(\boldsymbol{\rho},t)$ can still be derived from a (magnetic) scalar potential, which in the present context means that these components retain the properties of a genuine $\mathbf{H}$-field and that the magnetostatic boundary conditions can be applied to them. This is clearly not the case for the $z$-component, which must therefore be excluded from our analysis. A similar argument can be constructed for the $\mathbf{B}$-field.

In terms of these averaged fields, the boundary conditions can now be recast in the form
\begin{align}
    \lim_{x \rightarrow 0^-} B^{\mathrm{in}}_{\mathrm{ave},x}(\boldsymbol{\rho},t) &= \lim_{x \rightarrow 0^+} B^{\mathrm{in}}_{\mathrm{ave},x}(\boldsymbol{\rho},t),\label{Equation Boundary Condition B Appendix}\\
    \lim_{x \rightarrow 0^-} H^{\mathrm{in}}_{\mathrm{ave},y}(\boldsymbol{\rho},t)  &= \lim_{x \rightarrow 0^+} H^{\mathrm{in}}_{\mathrm{ave},y}(\boldsymbol{\rho},t). \label{Equation Boundary Condition H Appendix}
\end{align}
Recalling from Section~\ref{Section Solution inside magnetic insulator} that the average dipolar magnetizing field $\mathbf{H}_{\mathrm{d,ave}}^{\mathrm{in}}(\boldsymbol{\rho},t)$ and the average superconductor-response field $\mathbf{H}_{\mathrm{sc,ave}}^{\mathrm{in}}(\boldsymbol{\rho},t)$, upon removing the time-dependence $e^{-i\omega t}$, are given by
\begin{align}
\mathbf{H}^{\mathrm{in}}_{\mathrm{d},\mathrm{ave}}(\mathbf{k},\boldsymbol{\rho})
&=
\begin{cases}
\dfrac{e^{i\mathbf{k}\cdot \boldsymbol{\rho}}}{k^2}
\begin{pmatrix}
i k_x^2 [g(k)-1]\eta_0(\mathbf{k})\\
i k_x k_y[g(k)-1]\eta_0(\mathbf{k})\\
-k^2 g(k)
\end{pmatrix},
\hspace{1.4cm} (x<0),
\\[3mm]
\dfrac{e^{i\mathbf{k}\cdot \boldsymbol{\rho}}}{k^2}
\begin{pmatrix}
i k_x^2 [g(k)-1]\eta(\mathbf{k})\\
i k_x k_y[g(k)-1]\eta(\mathbf{k})\\
-k^2 g(k)
\end{pmatrix},
\hspace{1.6cm}  (x>0),
\end{cases}
\\[3mm]
\mathbf{H}^{\mathrm{in}}_{\mathrm{sc},\mathrm{ave}}(\mathbf{k},\boldsymbol{\rho})
&=
\begin{cases}
\mathbf{0}, \hspace{5.95cm} (x<0), \\
    -\frac{\omega_{\mathrm{sc}}(k)}{\gamma \mu_0 M_S}
\biggr[
\frac{k_x}{k}\eta(\mathbf{k}) + 1
\biggr]
e^{i\mathbf{k}\cdot\boldsymbol{\rho}}
\begin{pmatrix}
i k_x/k\\
i k_y/k\\
1
\end{pmatrix},
\hspace{0.7cm} (x>0),
\end{cases}
\end{align}
we find that Eqs.~\eqref{Equation Boundary Condition B Appendix} and \eqref{Equation Boundary Condition H Appendix} can be expressed as follows,
\begin{align}
    &\lim_{x \rightarrow 0^-} \bigg[\mathbf{B}_{\mathrm{d},\mathrm{ave}}^{\mathrm{in}}(\mathbf{k},\boldsymbol{\rho})+R_{12}\mathbf{B}_{\mathrm{d},\mathrm{ave}}^{\mathrm{in}}(\mathbf{k}_R,\boldsymbol{\rho})\bigg]_x = \lim_{x \rightarrow 0^+} T_{12}\bigg[\mathbf{B}_{\mathrm{d+sc},\mathrm{ave}}^{\mathrm{in}}(\mathbf{k}_T,\boldsymbol{\rho})\bigg]_x,\label{Equation 1 Appendix}\\
    &\lim_{x \rightarrow 0^-}\bigg[\mathbf{H}_{\mathrm{d},\mathrm{ave}}^{\mathrm{in}}(\mathbf{k},\boldsymbol{\rho})+R_{12} \mathbf{H}_{\mathrm{d},\mathrm{ave}}^{\mathrm{in}}(\mathbf{k}_R,\boldsymbol{\rho})\bigg]_y = \lim_{x \rightarrow 0^+} T_{12} \bigg[ \mathbf{H}_{\mathrm{d+sc},\mathrm{ave}}^{\mathrm{in}}(\mathbf{k}_T,\boldsymbol{\rho}) \bigg]_y,\label{Equation 2 Appendix}\\
    &\lim_{x \rightarrow 0^-} T_{21} \bigg[\mathbf{B}_{\mathrm{d},\mathrm{ave}}^{\mathrm{in}}(\mathbf{k}_T,\boldsymbol{\rho})\bigg]_x = \lim_{x \rightarrow 0^+} \bigg[\mathbf{B}_{\mathrm{d+sc},\mathrm{ave}}^{\mathrm{in}}(\mathbf{k},\boldsymbol{\rho}) + R_{21}\mathbf{B}_{\mathrm{d+sc},\mathrm{ave}}^{\mathrm{in}}(\mathbf{k}_R,\boldsymbol{\rho})\bigg]_x,\label{Equation 3 Appendix}\\
        &\lim_{x \rightarrow 0^-}T_{21}\bigg[ \mathbf{H}_{\mathrm{d},\mathrm{ave}}^{\mathrm{in}}(\mathbf{k}_T,\boldsymbol{\rho})\bigg]_y = \lim_{x \rightarrow 0^+} \bigg[  \mathbf{H}_{\mathrm{d+sc},\mathrm{ave}}^{\mathrm{in}}(\mathbf{k},\boldsymbol{\rho}) + R_{21} \mathbf{H}_{\mathrm{d+sc},\mathrm{ave}}^{\mathrm{in}}(\mathbf{k}_R,\boldsymbol{\rho}) \bigg]_y.\label{Equation 4 Appendix}
\end{align}
Here, we have defined $\mathbf{H}_{\mathrm{d+sc},\mathrm{ave}}^{\mathrm{in}}(\mathbf{k},\boldsymbol{\rho}) = \mathbf{H}_{\mathrm{d},\mathrm{ave}}^{\mathrm{in}}(\mathbf{k},\boldsymbol{\rho})+\mathbf{H}_{\mathrm{sc},\mathrm{ave}}^{\mathrm{in}}(\mathbf{k},\boldsymbol{\rho})$, with an analogous expression for $\mathbf{B}_{\mathrm{d+sc},\mathrm{ave}}^{\mathrm{in}}(\mathbf{k},\boldsymbol{\rho})$. The relationship between the $\mathbf{B}$-fields and their corresponding $\mathbf{H}$-fields is given by
\begin{align}
    \mathbf{B}_{\mathrm{d},\mathrm{ave}}^{\mathrm{in}}(\mathbf{k},\boldsymbol{\rho}) &= \mu_0\bigg[\mathbf{H}_{\mathrm{d},\mathrm{ave}}^{\mathrm{in}}(\mathbf{k},\boldsymbol{\rho})+ \mathbf{m}(\mathbf{k},\boldsymbol{\rho}) \bigg], \\
    \mathbf{B}_{\mathrm{sc},\mathrm{ave}}^{\mathrm{in}}(\mathbf{k},\boldsymbol{\rho}) &= \mu_0\mathbf{H}_{\mathrm{sc},\mathrm{ave}}^{\mathrm{in}}(\mathbf{k},\boldsymbol{\rho}),
\end{align}
where
\begin{align}
    \mathbf{m}(\mathbf{k},\boldsymbol{\rho})) &= 
     \begin{cases}
        e^{i\mathbf{k}\cdot\boldsymbol{\rho}}\,
        \begin{pmatrix}
            i \eta_0(\mathbf{k}) \\ 0 \\ 1
        \end{pmatrix}
        \hspace{2.2cm} (x < 0),\\ 
        e^{i\mathbf{k}\cdot\boldsymbol{\rho}}\,
        \begin{pmatrix}
            i \eta(\mathbf{k}) \\ 0 \\ 1
        \end{pmatrix}\hspace{2.2cm} (x > 0).
        \end{cases}
\end{align}
Now, upon introducing the \textit{dipolar length scale} $\Lambda_{\mathrm{d}}(\mathbf{k})$ and the \textit{superconductor length scale} $\Lambda_{\mathrm{sc}}(\mathbf{k})$,
\begin{align}
    \Lambda_{\mathrm{d}}(\mathbf{k}) &= \dfrac{k_x}{k^2} \bigg[ g(k) -1 \bigg],\\
    \Lambda_{\mathrm{sc}}(\mathbf{k}) &= \dfrac{\omega_{\mathrm{sc}}(k)}{\gamma \mu_0 M_S k} \bigg[ 1+ \dfrac{k_x}{k} \eta(\mathbf{k})\bigg],
\end{align}
we find that Eqs.~\eqref{Equation 1 Appendix}-\eqref{Equation 4 Appendix} can be expressed in matrix form as follows,
\begin{align}
    \begin{pmatrix}
        -\eta_0(\mathbf{k}_R) \Lambda_{\mathrm{d}}(\mathbf{k}_R) & \eta(\mathbf{k}_T)\Lambda_{\mathrm{d}}(\mathbf{k}_T)-\Lambda_{\mathrm{sc}}(\mathbf{k}_T)\\
        -\eta_0(\mathbf{k}_R) \bigg[\Lambda_{\mathrm{d}}(\mathbf{k}_R) k_{R,x} + 1 \bigg]& \eta(\mathbf{k}_T)\bigg[\Lambda_{\mathrm{d}}(\mathbf{k}_T)k_{T,x}+1\bigg]-\Lambda_{\mathrm{sc}}(\mathbf{k}_T)k_{T,x}
    \end{pmatrix}
    \begin{pmatrix}
        R_{12}\\
        T_{12}
    \end{pmatrix}
    &=
    \begin{pmatrix}
        \eta_0(\mathbf{k}) \Lambda_{\mathrm{d}}(\mathbf{k})\\
        \eta_0(\mathbf{k}) \bigg[\Lambda_{\mathrm{d}}(\mathbf{k}) k_x + 1 \bigg]
    \end{pmatrix}
    ,
\end{align}
\begin{align}
    \begin{pmatrix}
        -\eta(\mathbf{k}_R) \Lambda_{\mathrm{d}}(\mathbf{k}_R) +\Lambda_{\mathrm{sc}}(\mathbf{k}_R) & \eta_0(\mathbf{k}_T)\Lambda_{\mathrm{d}}(\mathbf{k}_T)\\
        -\eta(\mathbf{k}_R) \bigg[\Lambda_{\mathrm{d}}(\mathbf{k}_R) k_{R,x} + 1 \bigg]+\Lambda_{\mathrm{sc}}(\mathbf{k}_R)k_{R,x}& \eta_0(\mathbf{k}_T)\bigg[\Lambda_{\mathrm{d}}(\mathbf{k}_T)k_{T,x}+1\bigg]
    \end{pmatrix}
    \begin{pmatrix}
        R_{21}\\
        T_{21}
    \end{pmatrix}
    &=
    \begin{pmatrix}
        \eta(\mathbf{k}) \Lambda_{\mathrm{d}}(\mathbf{k})- \Lambda_{\mathrm{sc}}(\mathbf{k})\\
        \eta(\mathbf{k}) \bigg[\Lambda_{\mathrm{d}}(\mathbf{k}) k_x + 1 \bigg]- \Lambda_{\mathrm{sc}}(\mathbf{k})k_x
    \end{pmatrix}
    ,
\end{align}
where we have used the fact that $k_y$ is conserved. These systems of linear equations can be solved straightforwardly using Cramer's rule, and we immediately find
\begin{align}
    R_{12} & = \dfrac{\det
    \begin{pmatrix}
       \eta_0(\mathbf{k}) \Lambda_{\mathrm{d}}(\mathbf{k}) & \eta(\mathbf{k}_T)\Lambda_{\mathrm{d}}(\mathbf{k}_T)-\Lambda_{\mathrm{sc}}(\mathbf{k}_T)\\
       \eta_0(\mathbf{k}) \bigg[\Lambda_{\mathrm{d}}(\mathbf{k}) k_x + 1 \bigg]& \eta(\mathbf{k}_T)\bigg[\Lambda_{\mathrm{d}}(\mathbf{k}_T)k_{T,x}+1\bigg]-\Lambda_{\mathrm{sc}}(\mathbf{k}_T)k_{T,x}
    \end{pmatrix} 
    }{\det 
    \begin{pmatrix}
        -\eta_0(\mathbf{k}_R) \Lambda_{\mathrm{d}}(\mathbf{k}_R) & \eta(\mathbf{k}_T)\Lambda_{\mathrm{d}}(\mathbf{k}_T)-\Lambda_{\mathrm{sc}}(\mathbf{k}_T)\\
        -\eta_0(\mathbf{k}_R) \bigg[\Lambda_{\mathrm{d}}(\mathbf{k}_R) k_{R,x} + 1 \bigg]& \eta(\mathbf{k}_T)\bigg[\Lambda_{\mathrm{d}}(\mathbf{k}_T)k_{T,x}+1\bigg]-\Lambda_{\mathrm{sc}}(\mathbf{k}_T)k_{T,x}
    \end{pmatrix}
    }
    ,\\
    T_{12} &= \dfrac{\det
    \begin{pmatrix}
       -\eta_0(\mathbf{k}_R) \Lambda_{\mathrm{d}}(\mathbf{k}_R) & \eta_0(\mathbf{k}) \Lambda_{\mathrm{d}}(\mathbf{k})\\
       -\eta_0(\mathbf{k}_R) \bigg[\Lambda_{\mathrm{d}}(\mathbf{k}_R) k_{R,x} + 1 \bigg]& \eta_0(\mathbf{k}) \bigg[\Lambda_{\mathrm{d}}(\mathbf{k}) k_x + 1 \bigg]
    \end{pmatrix} 
    }{\det 
    \begin{pmatrix}
        -\eta_0(\mathbf{k}_R) \Lambda_{\mathrm{d}}(\mathbf{k}_R) & \eta(\mathbf{k}_T)\Lambda_{\mathrm{d}}(\mathbf{k}_T)-\Lambda_{\mathrm{sc}}(\mathbf{k}_T)\\
        -\eta_0(\mathbf{k}_R) \bigg[\Lambda_{\mathrm{d}}(\mathbf{k}_R) k_{R,x} + 1 \bigg]& \eta(\mathbf{k}_T)\bigg[\Lambda_{\mathrm{d}}(\mathbf{k}_T)k_{T,x}+1\bigg]-\Lambda_{\mathrm{sc}}(\mathbf{k}_T)k_{T,x}
    \end{pmatrix}
    }
    ,
\end{align}
and
\begin{align}
    R_{21} &= \dfrac{\det
    \begin{pmatrix}
        \eta(\mathbf{k}) \Lambda_{\mathrm{d}}(\mathbf{k})- \Lambda_{\mathrm{sc}}(\mathbf{k}) & \eta_0(\mathbf{k}_T)\Lambda_{\mathrm{d}}(\mathbf{k}_T)\\
        \eta(\mathbf{k}) \bigg[\Lambda_{\mathrm{d}}(\mathbf{k}) k_x + 1 \bigg]- \Lambda_{\mathrm{sc}}(\mathbf{k})k_x& \eta_0(\mathbf{k}_T)\bigg[\Lambda_{\mathrm{d}}(\mathbf{k}_T)k_{T,x}+1\bigg]
    \end{pmatrix}
    }{\det
    \begin{pmatrix}
        -\eta(\mathbf{k}_R) \Lambda_{\mathrm{d}}(\mathbf{k}_R) +\Lambda_{\mathrm{sc}}(\mathbf{k}_R) & \eta_0(\mathbf{k}_T)\Lambda_{\mathrm{d}}(\mathbf{k}_T)\\
        -\eta(\mathbf{k}_R) \bigg[\Lambda_{\mathrm{d}}(\mathbf{k}_R) k_{R,x} + 1 \bigg]+\Lambda_{\mathrm{sc}}(\mathbf{k}_R)k_{R,x}& \eta_0(\mathbf{k}_T)\bigg[\Lambda_{\mathrm{d}}(\mathbf{k}_T)k_{T,x}+1\bigg]
    \end{pmatrix}
    },
    \\
    T_{21} &= \dfrac{\det
    \begin{pmatrix}
        -\eta(\mathbf{k}_R) \Lambda_{\mathrm{d}}(\mathbf{k}_R) +\Lambda_{\mathrm{sc}}(\mathbf{k}_R) & \eta(\mathbf{k}) \Lambda_{\mathrm{d}}(\mathbf{k})- \Lambda_{\mathrm{sc}}(\mathbf{k})\\
        -\eta(\mathbf{k}_R) \bigg[\Lambda_{\mathrm{d}}(\mathbf{k}_R) k_{R,x} + 1 \bigg]+\Lambda_{\mathrm{sc}}(\mathbf{k}_R)k_{R,x}& \eta(\mathbf{k}) \bigg[\Lambda_{\mathrm{d}}(\mathbf{k}) k_x + 1 \bigg]- \Lambda_{\mathrm{sc}}(\mathbf{k})k_x
    \end{pmatrix}
    }{\det
    \begin{pmatrix}
        -\eta(\mathbf{k}_R) \Lambda_{\mathrm{d}}(\mathbf{k}_R) +\Lambda_{\mathrm{sc}}(\mathbf{k}_R) & \eta_0(\mathbf{k}_T)\Lambda_{\mathrm{d}}(\mathbf{k}_T)\\
        -\eta(\mathbf{k}_R) \bigg[\Lambda_{\mathrm{d}}(\mathbf{k}_R) k_{R,x} + 1 \bigg]+\Lambda_{\mathrm{sc}}(\mathbf{k}_R)k_{R,x}& \eta_0(\mathbf{k}_T)\bigg[\Lambda_{\mathrm{d}}(\mathbf{k}_T)k_{T,x}+1\bigg]
    \end{pmatrix}
    }
    .
\end{align}

Having obtained the general solution for the reflection and transmission coefficients, we now consider their behavior in the London dipolar limit. In said limit, we have
\begin{align}
    \omega_{\mathrm{sc}}(k) &\approx \gamma \mu_0 M_S \dfrac{k d}{2},\\
    g(k)-1 &\approx - \dfrac{k d}{2},
\end{align}
which yields the following approximate expressions for the dipolar and superconductor length scales:
\begin{align}
    \Lambda_{\mathrm{d}}(\mathbf{k)} &\approx -\dfrac{k_x}{k} \dfrac{d}{2},\\
    \Lambda_{\mathrm{sc}}(\mathbf{k)} &\approx \dfrac{d}{2} \bigg[ \dfrac{k_x}{k} \eta(\mathbf{k})+1 \bigg].
\end{align}
Multiplying both the numerators and denominators of the reflection and transmission coefficients by $k^2$, we retain only the terms up to linear order in $kd$ in both the numerators and denominators. This approximation is justified because, in the London dipolar limit, we have $kd \ll 1$. Finally, using the fact that $\lim_{kd \rightarrow 0}\,\eta(\mathbf{k}) = \lim_{kd \rightarrow 0}\,\eta_0(\mathbf{k}) = \xi$ and that $\mathbf{k}_R = (-k_x,k_y)$ in the FMI, we arrive at our final expressions for the reflection and transmission coefficients in the London dipolar limit:
\begin{align}
        R_{12}(\mathbf{k}) &= \dfrac{-\xi \bigg( \dfrac{2k_{T,x}}{k_T}-\dfrac{k_{x}}{k} \bigg)-1}{\xi \bigg( \dfrac{2k_{T,x}}{k_T}+\dfrac{k_{x}}{k} \bigg)+1}, \\
        T_{12}(\mathbf{k}) &= \dfrac{2\xi \dfrac{k_x}{k}}{\xi \bigg( \dfrac{2k_{T,x}}{k_T}+\dfrac{k_{x}}{k} \bigg)+1},
    \end{align}
    and
    \begin{align}
        R_{21}(\mathbf{k}) &= \dfrac{\xi \bigg( \dfrac{k_{T,x}}{k_T}-\dfrac{2k_{x}}{k} \bigg)-1}{\xi \bigg( \dfrac{2k_{R,x}}{k_R}-\dfrac{k_{T,x}}{k_T} \bigg)+1},\\
        T_{21}(\mathbf{k}) &= \dfrac{2\xi \bigg( \dfrac{k_{R,x}}{k_R}-\dfrac{k_{x}}{k} \bigg)}{\xi \bigg( \dfrac{2k_{R,x}}{k_R}-\dfrac{k_{T,x}}{k_T} \bigg)+1}. 
    \end{align}

\bibliographystyle{apsrev4-2}
\bibliography{biblio}
	
\end{document}